\documentclass{emulateapj}

\usepackage{graphicx}
\usepackage{amsmath}
\usepackage{natbib}
\usepackage{amssymb}

\usepackage[breaklinks,colorlinks,urlcolor=blue,citecolor=blue,linkcolor=blue]{hyperref}
\usepackage{hyperref}

\providecommand\natexlab[1]{#1}
\providecommand\JournalTitle[1]{#1}

\begin{document}

\newcommand{\be}{\begin{equation}}
\newcommand{\ee}{\end{equation}}

\newcommand{\BM}[1]{\textcolor{cyan}{#1}}

\title{A magnetar origin for the kilonova ejecta in GW170817}

\author{Brian D.~Metzger$^{1}$, Todd A.~Thompson$^{2}$, Eliot Quataert$^{3}$}
\altaffiltext{1}{Department of Physics and Columbia Astrophysics Laboratory, Columbia University, New York, NY 10027, USA.  email: bdm2129@columbia.edu}
\altaffiltext{2}{Department of Astronomy and Center for Cosmology \& Astro-Particle Physics, The Ohio State University, Columbus, Ohio 43210, USA}
\altaffiltext{3}{Departments of Physics and Astronomy, Theoretical Astrophysics Center, UC Berkeley, Berkeley, CA 94720, USA}

\begin{abstract}
The neutron star (NS) merger GW170817 was followed over several days by optical-wavelength  (``blue") kilonova (KN) emission likely powered by the radioactive decay of light $r$-process nuclei synthesized by ejecta with a low neutron abundance (electron fraction $Y_{e} \approx 0.25-0.35$).  While the composition and high velocities of the blue KN ejecta are consistent with shock-heated dynamical material, the large quantity is in tension with the results of numerical simulations.  We propose an alternative ejecta source: the neutrino-heated, magnetically-accelerated wind from the strongly-magnetized hypermassive NS (HMNS) remnant.  A rapidly-spinning HMNS with an ordered surface magnetic field of strength $B \approx 1-3\times 10^{14}$ G and lifetime $t_{\rm rem} \sim 0.1-1$ s can simultaneously explain the velocity, total mass, and electron fraction of the blue KN ejecta.  The inferred HMNS lifetime is close to its Alfv\'en crossing time, suggesting global magnetic torques could be responsible for bringing the HMNS into solid body rotation and instigating its gravitational collapse.  Different origins for the KN ejecta may be distinguished by their predictions for the emission in the first hours after the merger, when the luminosity is enhanced by heating from internal shocks; the latter are likely generic to any temporally-extended ejecta source (e.g. magnetar or accretion disk wind) and are not unique to the emergence of a relativistic jet.  The same shocks could mix and homogenizes the composition to a low but non-zero lanthanide mass fraction, $X_{\rm La} \approx 10^{-3}$, as advocated by some authors, but only if the mixing occurs after neutrons are consumed in the $r$-process on a timescale $\gtrsim 1$ s.  
\end{abstract}

\maketitle

\section{Source of the blue kilonova ejecta}

The gravitational wave chirp from the binary neutron star (NS) merger GW170817 \citep{LIGO+17DISCOVERY} was followed within seconds by a short burst of gamma-rays \citep{Goldstein+17,Savchenko+17,LIGO+17FERMI}.  Roughly half a day later, a luminous optical counterpart was discovered in the galaxy NGC 4993 at a distance of only $\approx 40$ Mpc (\citealt{Coulter+17,Soares-Santos+17,Arcavi+17,Diaz+17,Hu+17,Kilpatrick+17,Smartt+17,Drout+17,Evans+17,LIGO+17CAPSTONE,McCully+17,Buckley+17,Utsumi+17,Covino+17,Hu+17}).  The observed emission started out blue in color, with a featureless thermal spectrum that peaked at UV/optical frequencies (e.g.~\citealt{Nicholl+17,McCully+17,Evans+17}), before rapidly evolving over the course of a few days to become dominated by emission with a spectral peak in the near-infrared (NIR) \citep{Chornock+17,Pian+17,Tanvir+17}.  Simultaneous optical (e.g.~\citealt{Nicholl+17,Shappee+17}) and NIR \citep{Chornock+17} spectra around day 2.5 appear to demonstrate the presence of distinct optical and NIR emission components (however, see \citealt{Smartt+17}; \citealt{Waxman+17} for an alternative interpretation).

The properties of the optical/NIR emission agreed remarkably well with those predicted for ``kilonova" (KN) emission powered by the radioactive decay of $r$-process nuclei \citep{Li&Paczynski98,Metzger+10,Roberts+11,Barnes&Kasen13,Tanaka&Hotokezaka13,Grossman+14,Martin+15,Tanaka+17,Wollaeger+17,Fontes+17}; see \citealt{Fernandez&Metzger16}, \citealt{Metzger17} for reviews), a conclusion reached independently by several groups (e.g.~\citealt{Kasen+17,Drout+17,Tanaka+17b,Kasliwal+17,Murguia-Berthier+17,Waxman+17}).  The early-time blue emission is well-explained by radioactive material with a relatively low opacity (a ``blue" KN; \citealt{Metzger+10,Roberts+11}), as would be expected if the outer ejecta layers contain exclusively light $r$-process nuclei synthesized from matter with a relatively high electron fraction, $Y_{e} \gtrsim 0.25$ \citep{Metzger&Fernandez14}.  On the other hand, an upper bound of $Y_{e} \lesssim 0.3-0.35$ is needed to produce a radioactive decay rate consistent with the smooth decline in the bolometric light curve (e.g.~\citealt{Lippuner&Roberts15,Rosswog+17}).  The late NIR is instead well-explained by radioactive material with high line opacity, consistent with the presence of at least a moderate mass fraction $X_{\rm La} \gtrsim 10^{-2}$ of lanthanide or actinide nuclei (a so-called ``red" or ``purple" KN; \citealt{Kasen+13,Barnes&Kasen13,Tanaka&Hotokezaka13}).  Motivated by the theoretical expectation of distinct ejecta components with different compositions and opacities \citep{Metzger&Fernandez14,Perego+14}, a two-zone (``red"+``blue") KN model provides a reasonable fit to the photometric and spectroscopic data; these give best-fit values for the ejecta masses and mean velocities of the blue and red components of $M_{\rm blue} \approx 2\times 10^{-2}M_{\odot}, v_{\rm blue} \approx 0.25$ c and $M_{\rm red} \approx 0.05M_{\odot}, v_{\rm red} \approx 0.1$ c, respectively (e.g.~\citealt{Kasen+17,Cowperthwaite+17,Drout+17,Perego+17,Villar+17}).  

Several groups have made the case that the data is also consistent with a single ejecta component with a range of velocities and single low opacity \citep{Smartt+17,Tanaka+17b,Waxman+17}.  However, the thermal NIR continuum requires a higher line opacity in this wavelength range than supplied by light $r$-process nuclei (with exclusively d-shell valence electron shell structure), implicating the presence of some lanthanides/actinide nuclei in the inner ejecta layers \citep{Kasen+17}.  Synthesizing a small but non-zero lanthanide mass fraction (e.g.~$X_{\rm La}\approx 10^{-3}$) requires fine-tuning of the ejecta $Y_{e}$ distribution: the value of $X_{\rm La}$ increases from $\lesssim 10^{-4}$ to $\gtrsim 0.1$ over an extremely narrow range in $Y_{e}$ centered at $Y_{e} \approx 0.25$ \citep{Tanaka+17}.  Ejecta with a smoothly varying $Y_{e}$ outwards to higher velocities, with distinct lanthanide-rich and lanthanide-poor regions, therefore provides a more physical explanation for the observations (though we point out how delayed mixing of the ejecta might provide a loophole to the lanthanide fine-tuning argument in $\S\ref{sec:discussion}$).   

The total quantity of KN ejecta required to explain the observations $\gtrsim 0.06 M_{\odot}$ exceeds that of the dynamical ejecta for any general relativistic (GR) NS merger simulation published to date; furthermore, the velocity of the red/purple KN ejecta, $v_{\rm red} \approx 0.1$ c, is significantly less than predicted by simulations for the dynamical ejecta (which typically find $v \gtrsim 0.2-0.3$ c; e.g.~\citealt{Rosswog+99,Oechslin&Janka06,Bauswein+13,Hotokezaka+13,Sekiguchi+16,Dietrich+17,Bovard+17}).  The red KN ejecta is  instead more naturally explained as being an outflow from the remnant accretion torus, which is created around the central compact object following the merger (e.g.~\citealt{Metzger+08c,Metzger+09,Fernandez&Metzger13,Metzger&Fernandez14,Perego+14,Just+15,Martin+15,Wu+16,Lippuner+17,Siegel&Metzger17}).  Three-dimensional GR MHD simulation of the post-merger disk evolution demonstrate that $\approx 40\%$ of the initial mass of the torus will be unbound over a timescale of $\sim 1$ s post-merger at an average velocity of $v \approx 0.1$ c (e.g.~\citealt{Siegel&Metzger17,Siegel&Metzger17b}; see also \citealt{Metzger+08c}). 

If the merger end product is a hyper-massive NS (HMNS) remnant (as was likely the case for GW170817; e.g.~\citealt{Margalit&Metzger17,Granot+17,Bauswein+17,Perego+17,Rezzolla+17,Ruiz+17,Ma+17}), then the formation of a torus generically results from the outwards transport of angular momentum, which removes centrifugal support and results in the collapse of the HMNS to a black hole (e.g.~\citealt{Shibata&Taniguchi06}).  However, creating a torus of sufficient mass $\gtrsim 0.1 M_{\odot}$, as needed to supply the $M_{\rm red} \gtrsim 0.04 M_{\odot}$ of wind ejecta inferred for GW170817, requires a fairly stiff NS equation of state.  For instance, \citet{Radice+17} find that producing a torus of sufficient mass translates into a lower limit on the dimensionless NS tidal deformability parameter of $\tilde{\Lambda} \gtrsim 400$, consistent with the upper limit $\tilde{\Lambda} \lesssim 800$ obtained from the absence of detectable tidal losses on the inspiral gravitational waveform \citep{LIGO+17DISCOVERY}.  This bound $\tilde{\Lambda} \gtrsim 400$ translates into a rough lower limit on the radius of a 1.4$M_{\odot}$ NS of $\gtrsim 12$ km.

The source of the less neutron-rich ejecta responsible for the early blue KN is not so easily understood.  Table \ref{table:blueKN} summarizes several possible origins, with their successes and shortcomings.  While the high velocities $v_{\rm blue} \approx 0.2-0.3$ c and composition ($Y_{e} \gtrsim 0.25$) agree well with predictions for the shock-heated dynamical ejecta \citep{Oechslin&Janka06,Wanajo+14,Goriely+15,Sekiguchi+16,Radice+16}, the large quantity $M_{\rm blue} \approx 2\times 10^{-2}M_{\odot}$ is more challenging to explain.   \citet{Bauswein+13} perform a suite of equal-mass\footnote{For equal-mass mergers, most of the dynamical ejecta is shock-heated matter with high-$Y_{e}$ from the collision and not low-$Y_{e}$ tidal material.}  merger simulations for a range of different equations of states, finding $M_{\rm blue} \gtrsim 0.01M_{\odot}$ of dynamical ejecta only in cases where the NS radius is very small\footnote{Physically, a smaller NS radius results in greater shock-heated ejecta because more compact stars (of a fixed gravitational mass) collide at higher velocities.} $R_{\rm ns} \lesssim 11$ km \citep{Nicholl+17}, near the lower bound of $R_{\rm ns} \gtrsim 10.7$ km established by \citet{Bauswein+17} based on the inferred lack of a prompt collapse in GW170817.  As outlined above, such a small NS radius is also in tension with that needed to create the massive torus needed to explain the red KN \citep{Radice+17}, further challenging the dynamical explanation for the blue KN ejecta.  

The high velocity tail of the KN ejecta might not be an intrinsic property, but instead the result of shock-heating of an originally slower ejecta cloud by a relativistic jet created following some delay after the merger \citep{Bucciantini+12,Duffell+15,Gottlieb+18b,Gottlieb+18,Kasliwal+17,Bromberg+17,Piro&Kollmeier17}.  However, such a jet-accelerated (``cocoon") scenario still cannot account for the large quantity of high-$Y_{e}$ ejecta because a relativistic jet contributes negligible baryon mass.  The jet energy required to explain the kinetic energy of the blue KN ejecta $M_{\rm blue}v_{\rm blue}^{2}/2 \approx 10^{51}$ ergs furthermore exceeds the beaming-corrected kinetic energies of cosmological short gamma-ray bursts ($\approx 10^{49}-10^{50}$ ergs; \citealt{Nakar07,Berger14}) by a factor of $\sim 10-100$.

Here we propose an alternative source for the blue KN ejecta: a magnetized neutrino-irradiated wind, which emerges from the HMNS remnant over $\approx 0.1-1$ seconds prior to its collapse to a black hole.  Though a HMNS remnant was previously proposed as a potential ejecta source (e.g.~\citealt{Evans+17}), we emphasize the key role that strong magnetic fields play in endowing the wind with the high ejecta mass and velocity needed to explain the observations.  The properties of such proto-magnetar winds have been extensively explored previously in the context of gamma-ray bursts from core collapse supernovae (\citealt{Thompson+04,Bucciantini+06,Bucciantini+07,Bucciantini+08,Bucciantini+09,Metzger+07,Metzger+08a,Metzger+11,Vlasov+14,Vlasov+17}) and NS mergers (\citealt{Metzger+08b,Bucciantini+12}).  Bringing this theory to bear on GW170817, we find that a temporarily-stable magnetar remnant with a surface field strength $B \approx 1-3\times 10^{14}$ G can naturally provide the blue radioactive ejecta which lit up GW170817.  
  
\begin{table*}
\centering
\caption{Potential Sources of the Fast Blue KN  Ejecta in GW170817 \label{table:blueKN}}
\begin{tabular}{cccccc}
Ejecta Type & Quantity? & Velocity? & Electron Fraction? & References \\
\hline
Tidal Tail Dynamical & Maybe, if $M_{1}/M_{2} \lesssim 0.7^{\dagger}$ & \checkmark & {\bf Too Low} & e.g., 1, 2 &  \\
Shock-Heated Dynamical & Maybe, if $R_{\rm ns} \lesssim 11$ km$^{\ddagger}$ & \checkmark  & \checkmark  if NS long-lived & e.g., 3-5 &  \\
Accretion Disk Outflow & \checkmark if torus massive & {\bf Too Low} & \checkmark  if NS long-lived & e.g., 6-9&  \\
HMNS Neutrino-Driven Wind & {\bf Too Low} & {\bf Too Low} & {\bf Too High?} & e.g.,~11, 12\\
Magnetized HMNS Wind & \checkmark  if NS long-lived & \checkmark  & \checkmark  & e.g., 12,13&  \\
\hline \\
\end{tabular}
\\
$^{\dagger}$where here $M_{1}, M_{2}$ are the individual NS masses (see \citealt{Dietrich+17,Dietrich&Ujevic17,Gao+17}).
$^{\ddagger}$However, a small NS radius may be in tension with the creation of a large accretion disk needed to produce the red KN ejecta (\citealt{Radice+17}).  (1) \citealt{Rosswog+99}; (2) \citealt{Hotokezaka+13}; (3) \citealt{Bauswein+13}; (4) \citealt{Sekiguchi+16}; (5) \citealt{Radice+16}; (6) \citealt{Fernandez&Metzger13}; (7) \citealt{Perego+14}; (8) \citealt{Just+15}; (9) \citealt{Siegel&Metzger17}; (10) \citealt{Dessart+09}; (11) \citealt{Perego+14}; (12) \citealt{Metzger+08b}; (13) \citealt{Metzger+08a}
\end{table*}

\section{Neutrino-irradiated outflows from the HMNS Remnant}

The end product of a NS-NS merger is typically a HMNS remnant\footnote{
GW170817 could in principle have instead produced a long-lived supra-massive NS remnant, which was still temporarily supported against gravitational collapse by its solid body rotation even once differential rotation was removed; however, the prodigious rotational energy which must be deposited into its surroundings in order to collapse in this case would have been observationally conspicuous and thus is disfavored \citep{Margalit&Metzger17,Granot+17,Pooley+17,Margutti+17,Ma+17}.}, which is supported against gravitational collapse by its rapid and differential rotation (e.g.~\citealt{Shibata&Taniguchi06,Hanauske+17}).  Once its differential rotation has been removed (e.g. by gravitational waves or internal ``viscosity" resulting from various MHD instabilities) the HMNS collapses into a black hole; however, the timescale for this process, and thus the remnant lifetime $t_{\rm rem}$, is challenging to predict theoretically due to uncertainties in the supranuclear density equation of state and due to difficulties in numerically resolving all of the relevant physical processes over sufficiently long times (e.g.~\citealt{Siegel+13,Radice17,Shibata&Kiuchi17,Shibata+17}).  

As shown below, the mass-loss rate of the HMNS remnant over its first several seconds of cooling is so high as to preclude the formation of an ultra-relativistic jet (e.g.~\citealt{Metzger+08b,Murguia-Berthier+16}).  Thus, if the gamma-rays observed in coincidence with GW170817 (\citealt{Goldstein+17,Savchenko+17,LIGO+17FERMI}) are in some way associated with an off-axis$-$but otherwise typical$-$short GRB jet, then the formation of a black hole must have occurred prior to the arrival of the gamma-rays, setting an upper limit of $t_{\rm rem}\lesssim 1.7$ s on the HMNS lifetime.  On the other hand, as we discuss in $\S\ref{sec:discussion}$, the velocity of the magnetized HMNS wind could grow to become mildly relativistic over the neutrino cooling timescale of the remnant of a few seconds (\citealt{Metzger+08b}), in which case the gamma-rays could have been produced by the relativistic shock break-out created by internal collisions within the HMNS wind itself.

\subsection{Standard (unmagnetized) neutrino-driven wind}

One source of post-dynamical ejecta is a neutrino-driven wind from the hot NS remnant \citep{Dessart+09,Metzger&Fernandez14,Perego+14,Martin+15}, qualitatively similar to that expected from the hot proto-NS created by the core collapse of a massive star \citep{Duncan+86,Qian&Woosley96}.  The steady-state mass-loss rate due to neutrino heating from an unmagnetized NS is approximately given by  (\citealt{Qian&Woosley96,Thompson+01})
\begin{eqnarray}
\dot{M}_{\nu} &\simeq& 4\times 10^{-4}M_{\odot}\,{\rm s^{-1}}\left(\frac{L_{\nu}}{2\times 10^{52}\,{\rm ergs\,s^{-1}}}\right)^{5/3}\left(\frac{\epsilon_{\nu}}{15\,{\rm MeV}}\right)^{10/3}\nonumber \\
&&\left(\frac{M_{\rm ns}}{2.4M_{\odot}}\right)^{-2}\left(\frac{R_{\rm ns}}{15\,{\rm km}}\right)^{5/3},
\label{eq:Mdotnu}
\end{eqnarray} where we have combined the contributions of electron neutrinos and antineutrinos to the heating into a single product of the neutrino luminosity $L_{\nu}$ and mean neutrino energy $\epsilon_{\nu}$ defined by $L_{\nu}\epsilon_{\nu}^{2} \equiv L_{\nu_e}|\epsilon_{\nu_e}|^{2} + L_{\bar{\nu}_e}|\epsilon_{\bar{\nu}_e}|^{2}$.  The fiducial values used above are motivated by axisymmetric hydrodynamical simulations of the post-merger remnant including neutrino transport by \citet{Dessart+09}, who find $L_{\nu} \approx 2\times 10^{52}$ ergs s$^{-1}$ on timescales of $t \lesssim 30$ ms post-merger, with average neutrino energies in the range $\epsilon_{\nu} \approx 13-17$ MeV.  In agreement with the above, \citet{Dessart+09} found $\dot{M} \approx 10^{-3}-10^{-4}M_{\odot}$ s$^{-1}$ over the first $t \approx 20-100$ ms of evolution. 

Absent magnetic fields, the wind is accelerated entirely by thermal pressure and its  asymptotic velocity is approximately given by (\citealt{Thompson+01})
\be
v_{\nu} \approx 0.12\,{\rm c}\left(\frac{L_{\nu}}{2\times 10^{52}\,\rm ergs\,s^{-1}}\right)^{0.3},
\label{eq:vnu}
\ee

\begin{figure}
\centering
\includegraphics[width=0.5\textwidth]
{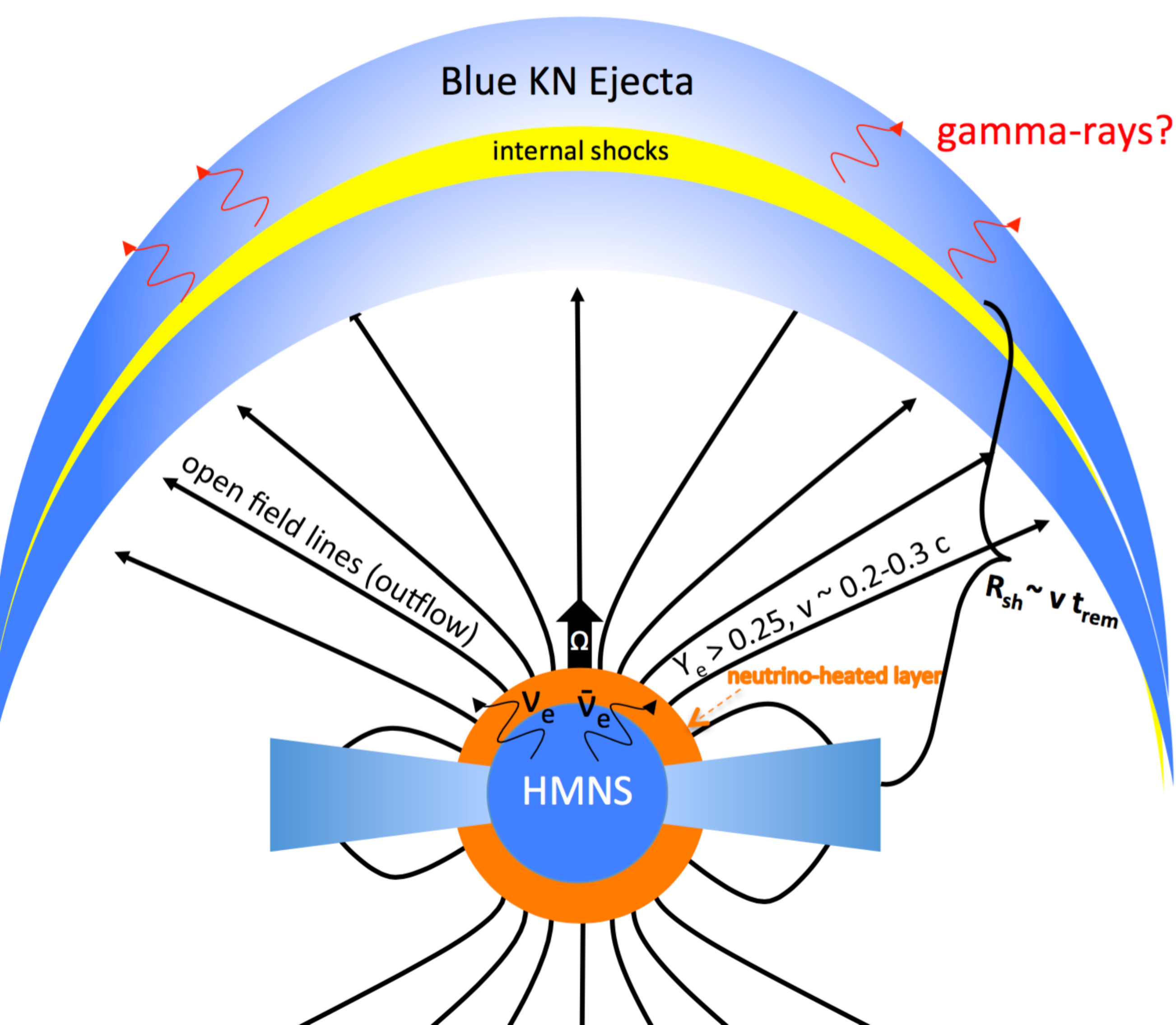}
\caption{\footnotesize Schematic diagram of the neutrino-irradiated wind from a magnetized HMNS.  Neutrinos from the HMNS heat matter in a narrow layer above the HMNS surface, feeding baryons onto open magnetic field lines at a rate which is substantially enhanced by magneto-centrifugal forces from the purely neutrino-driven mass-loss rate (e.g.~\citealt{Thompson+04,Metzger+07}).  Magnetic forces also accelerate the wind to a higher asymptotic velocity $v \approx v_{\rm B} \approx 0.2-0.3$ c (eq.~\ref{eq:vB}) than the purely neutrino-driven case $v_{\rm } \lesssim 0.1$ c (eq.~\ref{eq:vnu}), consistent with the blue KN ejecta.  Though blocked by the accretion disk directly in the equatorial plane, the outflow has its highest rate of mass-loss rate, kinetic energy flux, and velocity at low latitudes near the last closed field lines \citep{Vlasov+14}.  The wind velocity $\propto \sigma^{1/3} \propto B^{2/3}/\dot{M}^{1/3}$ may increase by a factor of $\sim 2$ over the HMNS lifetime (Fig.~\ref{fig:windevo}) as its mass loss rate $\dot{M}$ subsides, or its magnetic field $B$ is amplified, resulting in internal shocks on a radial scale $R_{\rm sh} \sim v t_{\rm rem} \sim 10^{10}(t_{\rm rem}/1{\rm s})$ cm, substantially larger than the wind launching point.  This late re-heating of the ejecta leads to brighter KN emission within the first few hours after the merger (Fig.~\ref{fig:KN}).  Relativistic break-out of the shocks as the magnetar wind becomes trans-relativistic on a similar timescale might also give rise to gamma-ray emission.   }
\label{fig:cartoon}
\end{figure}

The asymptotic electron fraction in the wind is set by the competition between the weak interactions $\nu_{e} + n \rightarrow p + e^{-}$ and  $\bar{\nu}_{e} + p \rightarrow n + e^{+}$; its equilibrium value is approximately given by \citep{Qian+93}
\be
Y_{e,\nu} \simeq \left(1 + \frac{L_{\bar{\nu}_{e}}}{L_{\nu_e}}\frac{\epsilon_{\bar{\nu}_e}-2\Delta + 1.2\Delta^{2}/\epsilon_{\bar{\nu}_e}}{\epsilon_{\nu_e}+2\Delta + 1.2\Delta^{2}/\epsilon_{\nu_e}}\right)^{-1} \approx 0.4-0.6,
\label{eq:Yenu}
\ee
where $\Delta \equiv (m_n-m_p)c^{2}$ is the neutron-proton mass difference.  This equilibrium is achieved in the wind because the gravitational binding energy of a nucleon on the surface of the HMNS of $GM_{\rm ns}m_p/R_{\rm ns}  \approx 200$ MeV greatly exceeds the mean neutrino energy $\epsilon_{\nu}$; each nucleon must therefore necessarily absorb several neutrinos in the process of escaping to infinity and the outflow is protonized as compared to the much lower value of $Y_{e} \lesssim 0.1$ at the base of the wind in the degenerate surface layers of the star.  In the final line, $Y_{e,\nu}$ is given as an range because the {\it difference} between the spectra of electron neutrinos and antineutrinos from the cooling merger remnant which enters the expression for $Y_{e,\nu}$ depends on the detailed transport processes in this environment and are thus theoretically uncertain (e.g.~\citealt{Roberts+12,MartinezPinedo+12} for recent work in the core collapse context).

The electron fraction of an unmagnetized PNS wind is sufficiently high $Y_{e} \gtrsim 0.4-0.5$ to synthesize exclusively Fe-group nuclei or light $r$-process nuclei with the low opacities needed to produce blue KN emission.  However, the quantity $\dot{M}_{\nu}t_{\rm rem} \lesssim 10^{-3}M_{\odot}$ and velocity $v_{\nu} \sim 0.1$ c of the neutrino-driven wind ejecta are too low compared to observations of GW170817.  The predicted composition may also be problematic; the radioactive energy input of $Y_{e} \approx Y_{e,\nu} \gtrsim 0.4$ matter is dominated by a few discrete nuclei \citep{Lippuner&Roberts15}, inconsistent with the observed smooth decay of the KN bolometric light curve (\citealt{Rosswog+17}).  Matter with lower $Y_{e}$ can be unbound by neutrino heating of the surrounding accretion disk (e.g.~\citealt{Metzger&Fernandez14,Perego+14,Martin+15}), but the velocity of this material $\lesssim 0.1$ c is also too low (Table \ref{table:blueKN}).

\subsection{Magnetized, neutrino-heated wind}
\label{sec:magnetar}

A standard neutrino-heated wind cannot explain the observed properties of the blue KN, but the prospects are better if the merger remnant possesses a strong magnetic field.  Due to the large orbital angular momentum of the initial binary, the remnant is necessarily rotating close to its mass-shedding limit, with a rotation period $P = 2\pi/\Omega \approx 0.8-1$ ms, where $\Omega$ is the angular rotation frequency.  The remnant is also highly magnetized, due to amplification of the magnetic field on small scales to $\gtrsim 10^{16}$ G by several instabilities (e.g.~Kelvin-Helmholtz, magneto-rotational) which tap into the free energy available in differential rotation (e.g.~\citealt{Price&Rosswog06,Siegel+13,Zrake&MacFadyen13,Kiuchi+15}).  As a part of this process, and the longer-term MHD evolution of its internal magnetic field (e.g.~\citealt{Braithwaite07}), the rapidly-spinning remnant could acquire a large-scale surface field, though its strength is likely to be weaker than the small-scale field.    

In the presence of rapid rotation and a strong ordered magnetic field, magneto-centrifugal forces accelerate matter outwards from the HMNS along the open field lines in addition to the thermal pressure from neutrino heating (Fig.~\ref{fig:cartoon}).  A magnetic field thus enhances the mass loss rate and velocity of the HMNS wind \citep{Thompson+04,Metzger+07}, in addition to reducing its electron fraction as compared to the equilibrium value obtain when the flow comes into equilibrium with the neutrinos, $Y_{e,\nu}$ (e.g.~\citealt{Metzger+08a}).   

A key property quantifying the dynamical importance of the magnetic field is the wind magnetization
\be
\sigma = \frac{\Phi_{\rm M}^{2}\Omega^{2}}{\dot{M}_{\rm tot}c^{3}} = \frac{B^{2}R_{\rm ns}^{4}f_{\rm open}\Omega^{2}}{\dot{M}c^{3}},
\label{eq:sigma}
\ee  
where $\Phi_{\rm M} = f_{\rm open} B R_{\rm ns}^{2}$ is the open magnetic flux per steradian leaving the NS surface, $B$ is the average surface magnetic field strength, $f_{\rm open}$ is the fraction of the NS surface threaded by open magnetic field lines, $\dot{M}_{\rm tot} = f_{\rm open}\dot{M}$ is the total mass loss rate, and $\dot{M}$ is the wind mass loss rate when $f_{\rm open} = 1$ limit (which in general will be substantially enhanced from the purely neutrino-driven value estimated in eq.~\ref{eq:Mdotnu}).  In what follows we assume the split-monopole magnetic field structure ($f_{\rm open} = 1$), which is a reasonable approximation if the magnetosphere is   continuously ``torn open" by latitudinal differential rotation \citep{Siegel+14}, neutrino heating of the atmosphere in the closed-zone region \citep{Thompson03,Komissarov&Barkov07,Thompson&udDoula17}, and by the compression of the nominally closed field zone by the ram pressure of the surrounding accretion disk \citep{Parfrey+16}.  However, our results can also be applied to the case $f_{\rm open} \ll 1$, as would characterize a more complex magnetic field structure, provided that the ratio $B^{2}/\dot{M} \propto f_{\rm open}^{-1}$ can be scaled-up accordingly to obtain the same value of $\sigma$ needed by observations.

Upon reaching the fast magnetosonic surface (outside of the light cylinder), the outflow achieves a radial four-velocity $v\gamma \simeq c\sigma^{1/3}$ \citep{Michel69}.  Winds with $\sigma \gg 1$ thus become ultra-relativistic, reaching a bulk Lorentz factor $\gamma \gg 1$ in the range $\sigma^{1/3} \lesssim \gamma \le \sigma$, depending on how efficiently additional magnetic energy initially carried out by Poynting flux is converted into kinetic energy outside of the fast surface.  By contrast, winds with $\sigma < 1$ attain sub-relativistic speeds given by\footnote{
This result can be understood to order-of-magnitude by noting that $v_{B} \approx R_{A}\Omega$, where $R_{A}$ is the Alfv\'en radius at which $B^{2}/8\pi \approx \rho v^2/2$, where $v$ and $\rho = \dot{M}/4\pi v r^{2}$ are the velocity and density of the wind at radius $r$ \citep{Thompson+04}.
} 
\begin{eqnarray}
v_{B} &\simeq& \sqrt{3}c\sigma^{1/3} =  \sqrt{3}\left(\frac{B^{2}R_{\rm ns}^{4}\Omega^{2}}{\dot{M}}\right)^{1/3} \nonumber \\
&\approx& 0.22{\rm c} \left(\frac{B}{10^{14}\,{\rm G}}\right)^{2/3}\left(\frac{\dot{M}}{0.01M_{\odot}\,{\rm s^{-1}}}\right)^{-1/3}\left(\frac{P}{\rm 1 ms}\right)^{-2/3},
\label{eq:vB}
\end{eqnarray}
where in the final line we have taken $R_{\rm ns} = 15$ km and the factor $\sqrt{3}$ accounts for the additional conversion of the wind Poynting flux (2/3 of its flow energy near the fast surface) into bulk kinetic energy at larger radii. 

\begin{figure*}
\centering
\includegraphics[width=1.0\textwidth]
{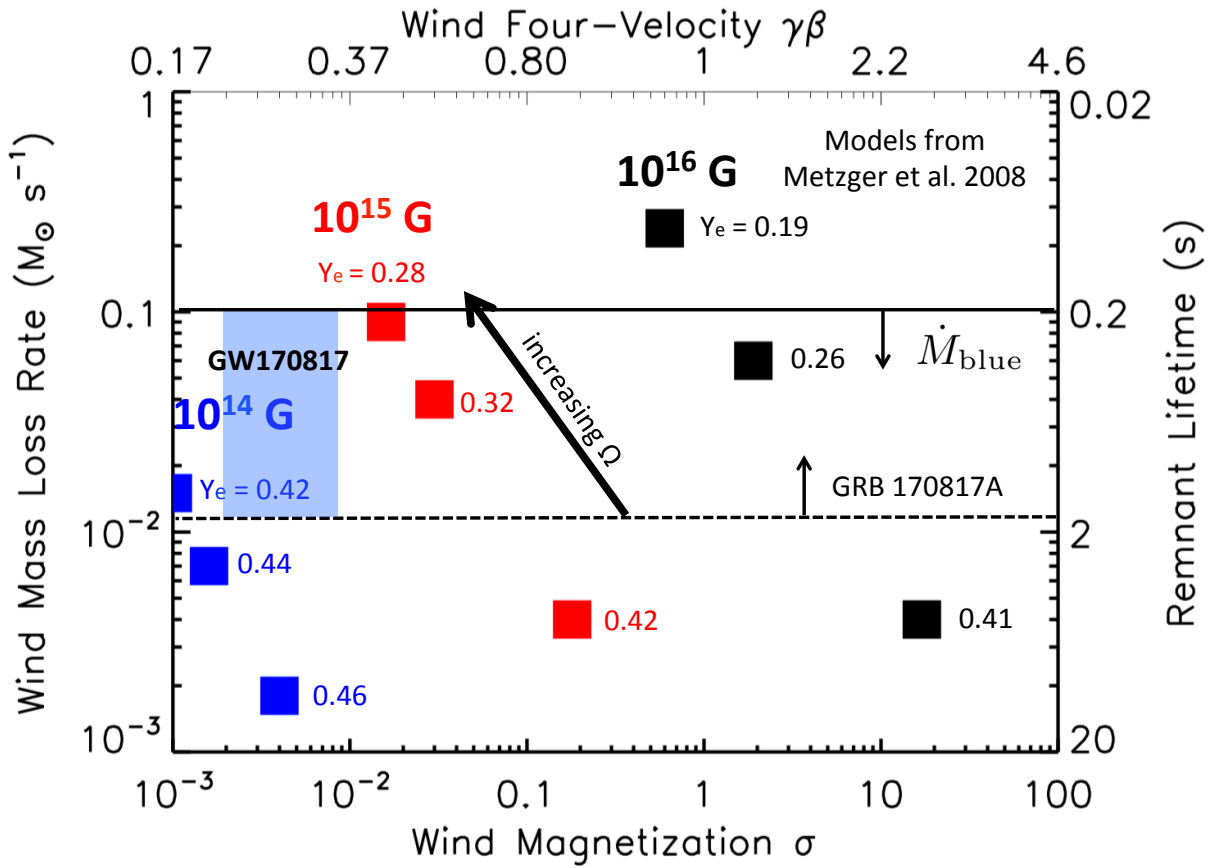}
\caption{\footnotesize Mass loss rate and magnetization (equivalently, asymptotic four-velocity $\beta\gamma$, where $\beta = v/c$ where $v$ is the wind velocity; top axis) from the steady-state, one-dimensional, neutrino-heated magneto-centrifugal wind solutions calculated by \citet{Metzger+08a}.  The solutions shown were calculated for a neutrino luminosity $L_{\nu} = L_{\nu_e} + L_{\bar{\nu}_e} \approx 1.4\times 10^{52}$ ergs s$^{-1}$ and mean neutrino energies $\epsilon_{\nu_e} = 11$ MeV, $\epsilon_{\bar{\nu}_e}$ = 14 MeV, similar to those from the HMNS remnant on timescales $\lesssim 1$ s after the merger \citep{Dessart+09}.  The right axis shows the remnant lifetime $t_{\rm rem}$ required to produce the total blue ejecta mass $M_{\rm blue} \approx 2\times 10^{-2}M_{\odot}$ of GW170817 given a constant mass loss rate $\dot{M}$ shown on the left axis. Each square shows a distinct steady-state solution, calculated for a different values of the surface magnetic field strength ($B = 10^{14}, 10^{15}, 10^{16}$ G, denoted by blue, red, and black colors, respectively) and for three rotation rates ($\Omega = 6000, 7000, 8000$ rad s$^{-1}$, from bottom to top) in the range necessary for a rapidly-spinning HMNS.   The asymptotic electron fraction of the wind is also marked next to each solution; this value is sufficiently high $Y_{e} \gtrsim 0.25$ to avoid significant lanthanide production for $\dot{M} \lesssim \dot{M}_{\rm blue}$ (eq.~\ref{eq:Mdotblue}; solid black line).  A magnetar remnant with $B \sim 1-3\times 10^{14}$ G, $\Omega \approx 9000$ Hz (P $\approx 0.8$ ms), and lifetime $t_{\rm rem} \approx 0.2-2$ can simultaneously explain the velocity $v_{\rm blue} \approx 0.3$ c, total mass $M_{\rm blue} \approx 2\times 10^{-2}M_{\odot}$, and composition $Y_{e} \approx 0.25-0.4$ of the blue KN emission from GW170817; this concordance region is shaded blue.  }
\label{fig:wind}
\end{figure*}

Figure \ref{fig:wind} shows the values of $\sigma$ (or, equivalently, asymptotic four-velocity; top axis) and $\dot{M}$ from a suite of steady-state, one-dimensional, neutrino-heated, magneto-centrifugal wind solutions calculated by \citet{Metzger+08a} for an assumed neutrino luminosity $L_{\nu} \approx 1.6\times 10^{52}$ ergs s$^{-1}$, similar to that from the hot post-merger remnant at early times $\sim 0.1-1$ s after the merger \citep{Dessart+09}.  Solutions are shown for different values of the surface magnetic field strength $B = 10^{14},10^{15},10^{16}$ G (denoted by different colors, from left to right) and several rotation rates $\Omega = 6000, 7000, 8000$ rad s$^{-1}$ ($P = 1.05, 0.90, 0.79$ ms; bottom to top) in the range expected for the post-merger remnant.  

The strong magnetic field increases both the kinetic energy and mass-loss rate of the wind.  $\dot{M}$ is increased by a factor of $\gtrsim 100-1000$ for $B \sim 10^{14}-10^{16}$ G as compared to the unmagnetized value given in eq.~\ref{eq:Mdotnu}.  At fixed $B$, $\dot{M}$ also rapidly increases with faster rotation, while the outflow velocity slightly {\it decreases} for larger $\Omega$ because of the $v_{\rm B} \propto \dot{M}^{-1/3}$ dependence.

The additional magneto-centrifugal acceleration reduces the time matter spends in the neutrino-heated region and reduces the final electron fraction $Y_{e}$ of the outflow compared to the neutrino-driven equilibrium value $Y_{e,\nu}$ (eq.~\ref{eq:Yenu}), as marked next to each solution in Fig.~\ref{fig:wind}.  While a relatively weak magnetic field ($10^{14}$ G) and slow rotation results in a value $Y_{e} \approx Y_{e,\nu} \simeq 0.46$ similar to the unmagnetized case, the more strongly-magnetized and rapidly-rotating cases have $Y_{e} \ll Y_{e,\nu}$.  In these cases the magnetic field accelerates matter away from the NS surface so rapidly that electron neutrinos have insufficient time to convert the initially neutron-rich composition at the electron-degenerate base of the wind back into one with a nearly equal number of neutrons and protons by much larger radii (where the $r$-process itself takes place).

Magneto-centrifugal enhancement of $\dot{M}$  is crucially tied to the conditions required for $Y_{e} \ll Y_{e,\nu}$, as both are related to the timescale matter has to absorb neutrinos near the base of the wind.  In the limit that the magnetic field dominates the wind dynamics in the neutrino heating region, an outflow which is sufficiently neutron-rich ($Y_{e}\lesssim 0.25$) to synthesize lathanide nuclei is necessarily accompanied by a minimum mass loss rate (\citealt{Metzger+08a}; their eq.~20)
\begin{eqnarray}
&&\dot{M}_{\rm blue} \approx 20\left(\frac{GM_{\rm ns}m_p}{R_{\rm ns}\epsilon_{\nu}}\right)\dot{M}_{\nu} \approx  0.1M_{\odot}\,{\rm s^{-1}}\times \nonumber \\
&&\left(\frac{L_{\nu}}{2\times 10^{52}\,{\rm ergs\,s^{-1}}}\right)^{5/3}\left(\frac{\epsilon_{\nu}}{15\,{\rm MeV}}\right)^{7/3}\left(\frac{M_{\rm ns}}{2.4M_{\odot}}\right)^{-1}.
\label{eq:Mdotblue}
\end{eqnarray}
The factor $(GM_{\rm ns}m_p/R_{\rm ns}\epsilon_{\nu})$ accounts for the additional energy per nucleon supplied which must be by the magnetic acceleration, above the normal value from neutrino heating acting alone, in order to obtain $Y_{e} \lesssim Y_{e,\nu}$.  The numerically-calibrated pre-factor of 20 is that needed to reduce $Y_{e}$ further to $\lesssim 0.25$.  Thus, one can place a strict upper limit on the quantity of blue KN ejecta ($Y_{e} \gtrsim 0.25$) of 
\be
M_{\rm blue}^{\rm max} \sim \dot{M}_{\rm blue}t_{\rm rem} \sim 10^{-2}(t_{\rm rem}/0.1{\rm\, s})M_{\odot},
\label{eq:Mbluemax}
\ee
where we have taken $L_{\nu} \approx 10^{52}$ ergs s$^{-1}$ as the roughly constant neutrino luminosity of the remnant on timescales of $t_{\rm rem} \lesssim 1$ s (e.g.~\citealt{Dessart+09}). 

The 1D solutions of \citet{Metzger+08a} were calculated along flow lines which emerge along the equatorial plane, near the last closed field line.  Because these low-latitude lines cover a large solid angle and carry most of the total mass flux of the wind, the results are similar to the total mass-loss rate and mass-averaged value of the outflow $Y_{e}$ integrated across the full open magnetosphere \citep{Vlasov+14,Vlasov+17}.  Nevertheless, the wind velocity and composition do show moderate variation with polar latitude $\theta$ of the outflow launching point; for instance, \citet{Vlasov+14} found that for $P \approx 1$ ms and $L_{\nu} \approx 2\times 10^{52}$ ergs s$^{-1}$, $Y_{e}$ varied from $\approx $ 0.40 to $\approx$ 0.32 as $\theta$ increased from $\lesssim 0.35$ to $ \approx 0.5$ (flow lines which carry $\approx 20\%$ and $\approx 80\%$ of the total mass loss, respectively).



\subsubsection{Application to the blue KN of GW170817}

On the timescale $t \lesssim t_{\rm rem} \lesssim 1$ s before the HMNS remnant collapses into a black hole, neither its rotation rate nor its neutrino luminosity will evolve substantially.   The average rotation frequency $\Omega$ cannot change substantially because only a moderate loss of rotational support would trigger gravitational collapse, while the neutrino luminosity decreases by a factor of $\lesssim$ 2 over the first second of the Kelvin-Helmholtz cooling evolution; e.g.~\citealt{Pons+99}).  Our analytic and numerical results at fixed $\Omega$ can be therefore used to approximate what type of magnetar remnant, if any, can explain all three key properties (mass, velocity, composition) the blue KN from GW170817 (Fig.~\ref{fig:wind}).  

For a wind duration equal to the magnetar lifetime $t_{\rm rem}$, producing the observed quantity of ejecta $M_{\rm blue} \approx 2\times 10^{-2}M_{\odot}$ (e.g.~\citealt{Cowperthwaite+17,Nicholl+17,Villar+17,Perego+17}) requires an average wind mass loss rate 
\be
\dot{M} \approx 2\times 10^{-2}(t_{\rm rem}/1\,{\rm s})^{-1}M_{\odot}\,s^{-1}
\ee
According to eq.~\ref{eq:vB}, explaining the velocity of the ejecta $v_{\rm blue} \approx 0.2-0.3\,{\rm c} = v_{\rm B}$ requires a wind magnetization $\sigma \approx 0.002-0.005$, which from eq.~(\ref{eq:sigma}) requires a surface magnetic field strength of (for $\Omega = 8000$ rad s$^{-1}$, $R = 15$ km)
\be
B \approx 1-2\times 10^{14}(t_{\rm rem}/1\,{\rm s})^{-1/2}\,{\rm G},
\ee
or $B \sim 1-3\times 10^{14}$ G for $t_{\rm rem} \sim 0.1-1$ s.  Extrapolating between the $B = 10^{14}$ G and $10^{15}$ G solutions in Fig.~\ref{fig:wind} shows that winds with mass-loss rates corresponding to $t_{\rm rem} \lesssim 1$ s indeed have electron fractions $Y_{e} \approx 0.25-0.4$ in agreement with the range required to power the blue KN.  This ``concordance" region is shaded blue in Fig.~\ref{fig:wind}.


\section{The first few hours }
\label{sec:earlytime}

Following its discovery at $t \approx 11$ hours, the first few days of optical emission following GW170817 is well-explained by KN emission powered by the decay of heavy $r$-process nuclei.  This agreement is independent of the precise origin of the ejecta because, by this late time, most of the initial thermal energy released during the ejection process on timescales $t_{0} \lesssim 1$ s has been degraded by adiabatic expansion, such that the observed luminosity is dominated by the radioactive decay of isotopes with half-lives $\sim t$.  However, different sources of blue KN ejecta will make quantitatively distinct predictions for the emission at earlier times, within the first few hours after the merger.  Though not available for GW170817, such early observations may be possible for future GW-detected mergers, such as those which occur above the North American continent at night (e.g.~\citealt{Kasliwal&Nissanke14}).  Early UV follow-up would also be possible with a dedicated wide-field satellite, such as the proposed ULTRASAT mission (e.g.~\citealt{Waxman+17})

A key distinguishing feature of different ejecta sources is the timescale over which the mass loss occurs.  Dynamical ejecta emerges almost immediately, within $\lesssim 1-10$ ms following the merger, and freely expands thereafter without additional heating other than from radioactivity.  By contrast, outflows from a magnetar or accretion disk emerge over longer timescales $\Delta t \sim 0.1-1$ s, in which case outflow variability acting over the same characteristic timescale would create internal shocks that inject additional thermal energy on radial scales $\sim \Delta t \cdot v_{\rm blue} \sim  10^{9}-10^{10}$ cm which greatly exceed\footnote{Heating at large radii might also be possible in rare instances by collision of the ejecta with a companion star in an evolved triple system \citep{Chang&Murray18}.} those of the central engine $\lesssim 10-100$ km (e.g.~\citealt{Beloborodov14}).  

As the HMNS cools following the merger, approximately over its Kelvin-Helmholtz diffusion timescale $\Delta t \sim 1$ s (e.g.~\citealt{Pons+99}), a factor of $\approx 2$ decrease in its neutrino luminosity would reduce the wind mass-loss rate by a factor of $\approx 3$ (eq.~\ref{eq:Mdotnu}) and increase the wind outflow speed $ v_{\rm B} \propto \dot{M}^{-1/3}$ by a factor of $\sim 1.5$ (eq.~\ref{eq:vB}); growth of the magnetic field strength from a dynamo, or emergence of magnetic flux from the interior, would also increase the wind speed $v_{\rm B} \propto B^{2/3}$.  The rising wind velocity will result in the development of internal shocks, heating the ejecta as discussed above.

\begin{figure}
\centering
\includegraphics[width=0.5\textwidth]
{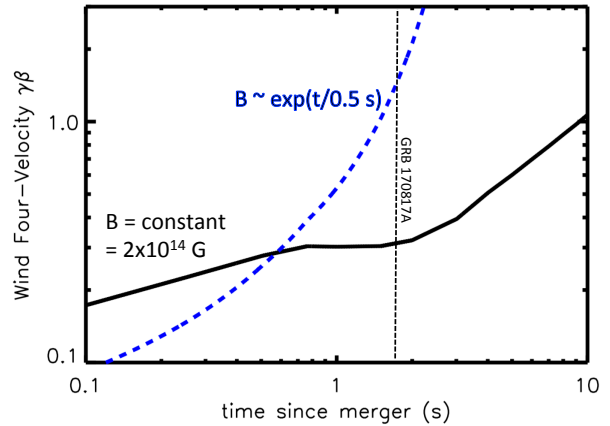}
\caption{\footnotesize Time evolution of the 4-velocity $\gamma \beta$ of the hyper-massive magnetar wind, calculated assuming that the wind mass-loss rate evolves as $\dot{M} \propto L_{\nu}^{5/3}\epsilon_{\nu}^{10/3}$, where the cooling evolution of the neutrino luminosity $L_{\nu}(t)$ and mean neutrino energy $\epsilon_{\nu}(t)$ are taken from the $2.0M_{\odot}$ proto-NS model of \citet{Pons+99}.  We separately show a case in which the magnetic field strength is held constant in time at $B \approx 2\times 10^{14}$ G (black solid line) and one in which the field strength grows exponentially on a timescale of $0.5$ s (dashed blue line).  In both cases the wind velocity over the first $\sim 0.1-1$ s are in the range $\sim 0.1-0.3$ c needed to explain the blue KN ejecta, but in the latter case the wind becomes trans-relativistic ($\gamma \beta \gtrsim 1$) by the onset of GRB 170817A, in which case relativistic break-out of internal shocks within the wind \citep{Nakar&Sari12} could serve as a potential origin for the gamma-rays.}
\label{fig:windevo}
\end{figure}

A crude but illustrative estimate of the time evolution of the wind velocity is shown in Fig.~\ref{fig:windevo}, separately for a case in which the HMNS magnetic field is held constant in time, and when the field is assumed to grow exponentially on a timescale of $\approx 0.5$ s.  The wind 4-velocity is calculated from eq.~\ref{eq:vB} based on the evolution of $\dot{M} \propto L_{\nu}^{5/3}\epsilon_{\nu}^{10/3}$ (eq.~\ref{eq:Mdotnu}), where $L_{\nu}(t)$ and $\epsilon_{\nu}(t)$ are taken from the proto-NS cooling calculations of \citet{Pons+99} (which should qualitatively capture the cooling evolution of the post-merger remnant).  In the case of temporally-strengthening field, the internal shocks in the magnetar wind could grow to trans-relativistic velocities $\gamma \beta \gtrsim 1$ on a timescale of couple seconds, if the HMNS has not already collapsed by this time.  Relativistic shock break-out (e.g.~\citealt{Nakar&Sari12}) from this interaction might provide an alternative source for the gamma-ray emission from GW170817, which does not necessarily implicate black hole formation or the creation of an ultra-relativistic jet on this timescale.  Future hydrodynamical modeling with radiative transfer is required to assess whether the internal shocks generated within the magnetar wind can explain in detail the properties (mean energy, luminosity, duration) of the gamma-ray emission. 

\begin{figure}
\centering
\includegraphics[width=0.5\textwidth]
{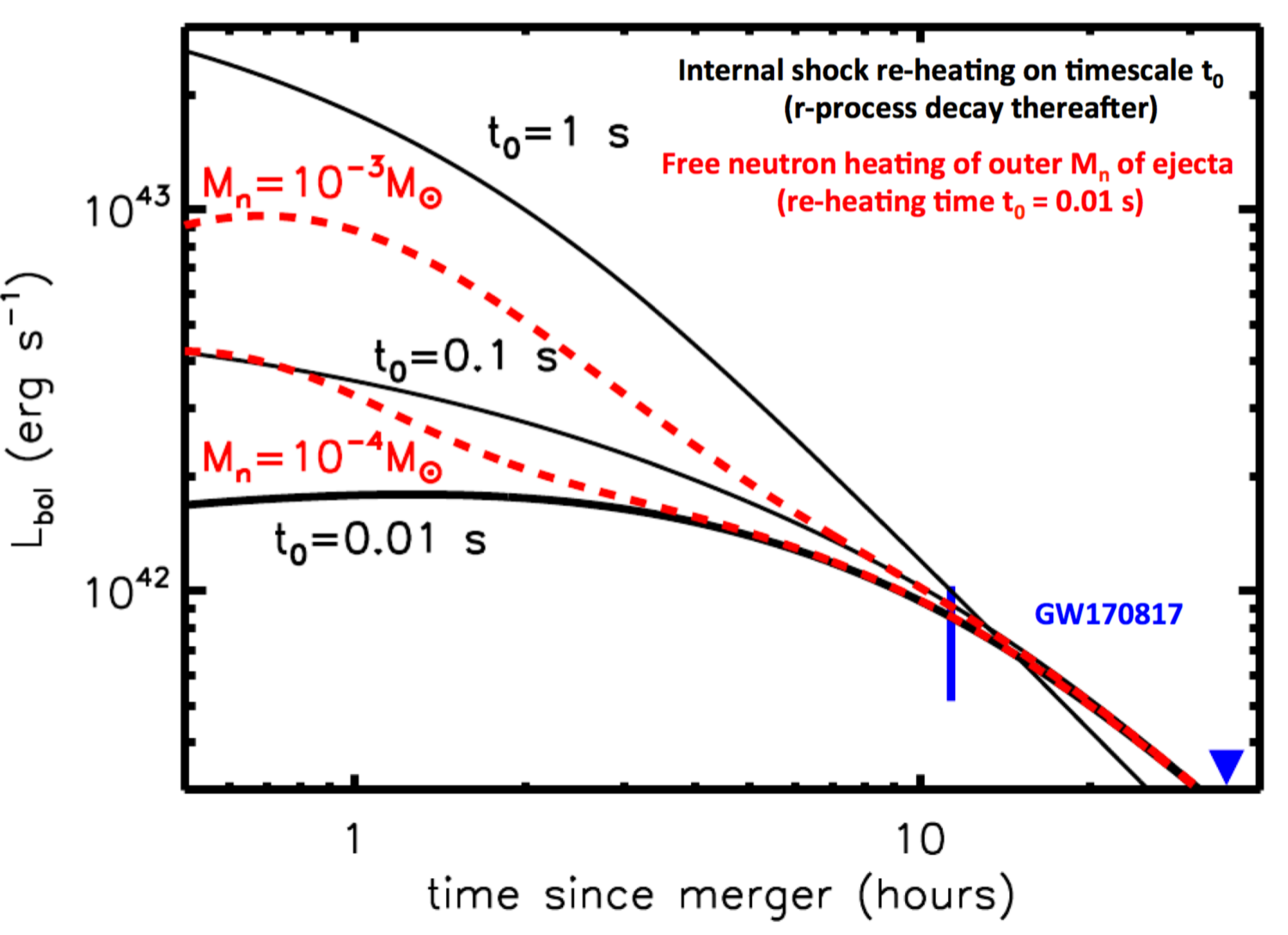}
\caption{\footnotesize Bolometric light curve of the blue KN emission during the first few hours after the merger, calculated for several models which reproduce the measured luminosity $L_{\rm bol} \approx 10^{42}$ ergs s$^{-1}$ of GW170817 at $t \approx 11$ hours (blue uncertainty bar; e.g.~\citealt{Cowperthwaite+17,Arcavi+17,Drout+17}).  Black solid lines show models for different assumptions about the timescale $t_{0} = 0.01, 0.1, 1$ s following ejection at which the matter was last thermalized, i.e.~possessed an internal thermal energy comparable to its asymptotic kinetic energy (for $t \ge t_0$, the heating is solely due to $r$-process radioactivity).  A small value of $t_{0} \lesssim 0.01$ s corresponds to a dynamical ejecta origin with no additional heating, while a large value of $t_{0} \gtrsim 0.1-1$ s represents the case of a long-lived engine (magnetar wind or accretion disk outflow) where the outflow properties vary over timescales $\approx t_{0}$.  We have assumed KN parameters (see Appendix): $\beta = 3$, $v_{0} = 0.25c$, $M_{\rm tot} = 0.025M_{\odot}$, $\kappa = 0.5$ cm$^{2}$ g$^{-1}$ (except for the $t_{0} = 1$ s case, for which $M_{\rm tot} = 0.015M_{\odot}$).  Red dashed lines show models with $t_{0} = 0.01$ s in which the outer $10^{-3}M_{\odot}$ or $10^{-4}M_{\odot}$ of the ejecta contain free neutrons instead of $r$-process nuclei \citep{Metzger+15}, showing how the enhanced luminosity from neutron decay (the ``neutron precursor") is degenerate with large thermalization times. }
\label{fig:KN}
\end{figure}

To illustrate the influence of radially extended heating on the early-time KN emission, Fig.~\ref{fig:KN} shows several KN models which fit the bolometric luminosity $L_{\rm bol} \approx 10^{42}$ ergs s$^{-1}$ measured at $t \approx 11$ hours (see Appendix for details) yet do not overproduce the luminosity at 1.5 days (at which time the slower inner layers which produce the red/purple KN are beginning to contribute).  Solid black lines show models for ejecta for different assumptions about the time $t_{0} = 0.01, 0.1, 1$ s after the merger at which the thermal energy of each layer of the ejecta was last comparable to its asymptotic kinetic energy (at times $t \ge t_0$, the only heating is by the decay of $r$-process nuclei).  As discussed above, this timescale can be roughly associated with the timescale $\Delta t$ over which a significant fraction of the mass loss is occurring, e.g. the HMNS lifetime $t_{\rm rem}$ in the case of a magnetar wind or the viscous timescale in the case of a jet or accretion disk outflow (e.g.~\citealt{Fernandez&Metzger13}).  Fig.~\ref{fig:KN} shows that the luminosity at one hour post-merger is $\sim 3-10$ times higher if the mass loss emerges and continues to self-interact through internal shocks over an extended timescale of $t_{0} \sim 0.1-1$ s (e.g. magnetar or accretion disk wind) than if the ejecta is dynamical in origin and freely expands from the merger site ($t_{0} \lesssim 0.01$ s). 

 Additional thermal energy could be deposited into the expanding matter also by the passage of a relativistic jet through the blue ejecta, creating an expanding tail of high velocity shock-heated matter (e.g.~\citealt{Gottlieb+18,Gottlieb+18b,Bromberg+17,Kasliwal+17,Piro&Kollmeier17}).  However, reasonable constraints can be placed on the magnitude of this effect based on the observed kinetic energies of on-axis cosmological short duration gamma-ray bursts.  The isotropic kinetic energies of short GRB jets as inferred from their non-thermal afterglows are typically $E_{\rm iso,K} \approx 10^{50}-10^{51}$ ergs (\citealt{Nakar07,Berger14,Fong+15}).  The local ($z = 0$) rate of binary NS mergers as measured by Advanced LIGO after GW170817 is $\mathcal{R}_{\rm BNS} = 1540^{+3200}_{-1220}$ Gpc$^{-3}$ yr$^{-1}$, while the observed rate of short GRBs is $\mathcal{R}_{\rm SGRB} \approx 2-6$ Gpc$^{-3}$ yr$^{-1}$ (\citealt{Wanderman&Piran15}).  This places an upper limit on the beaming fraction of the emission of $f_{\rm b} \lesssim 0.02 $, which in turn implies a beaming-corrected GRB jet energy of $\lesssim 10^{49}$ ergs which is 2 orders of magnitude smaller than the measured kinetic energy of the blue KN ejecta of $\sim 10^{51}$ ergs.  Thus, unless the jet cocoon contains $\gtrsim 10-100$ times more energy than is typical of the ultra-relativistic GRB itself, cocoon heating contributes insignificantly to the early-time optical emission.  This argument would not be valid if cosmological short GRBs do not originate from NS-NS mergers.

The early KN emission can also be enhanced due to the radioactive decay of free neutrons in the outermost layers of the ejecta (\citealt{Metzger+15}; see also \citealt{Kulkarni05}).  Free neutrons may be present in the outermost $\sim 1\%$ of the ejecta if these layers are expanding sufficiently rapidly for neutron-capture reactions to freeze-out prior to completion of the $r$-process (as occurs on a timescale of $\sim 1$ second; \citealt{Bauswein+13}).  The anomalously long half-life of a free neutron $\tau_{\rm n} \approx 10$ minutes, as compared to a typical $r$-process isotope ($\tau \sim 1$ s), enables free neutron decay to dominate $r$-process heating during the first several hours of emission.  Figure \ref{fig:KN} shows that the presence of free neutrons in the outermost $\sim 10^{-4}-10^{-3}M_{\odot}$ of the expanding material can mimick to some extent the effects of a long heating timescale $t_{\rm rem} \gtrsim 0.1-1$ s, including a cocoon.

\section{Discussion}
\label{sec:discussion}

While the presence of luminous optical (blue) KN emission from the first NS merger was not a surprise (e.g.~\citealt{Metzger+10}), the large ejecta mass, high velocity, and relatively neutron-poor composition are in tension with traditionally-considered ejecta sources (Table \ref{table:blueKN}).  Instead, we have proposed that the blue KN ejecta in GW170817 originates from the neutrino-irradiated, magneto-centrifugally-driven outflow from a temporarily-stable rapidly-rotating HMNS remnant (Fig.~\ref{fig:cartoon}).  Fig.~\ref{fig:wind} shows that such a millisecond magnetar wind can simultaneously explain the large ejecta mass (due to magneto-centrifugal enhancement of the neutrino-driven mass-loss), high velocity (magneto-centrifugal enhancement of the wind acceleration), and electron fraction (from moderate, but not excessive, neutrino irradiation near the wind launching point), given only two free parameters: the surface magnetic field strength $B \sim 1-3\times 10^{14}$ G  and HMNS lifetime $t_{\rm rem} \sim 0.1-1$ s.  

The HMNS generally collapses to a black hole once its centrifugal support from internal differential rotation is removed by the outwards transport of angular momentum (e.g.~\citealt{Kaplan+14}).  While this process has been modeled as local ``effective viscosity" (e.g.~\citealt{Radice17,Shibata+17}), angular momentum redistribution could in principle be dominated by torques from large-scale magnetic fields, even if they are weaker in magnitude than the smaller-scale turbulent ones.  In this case, solid-body rotation will be established roughly on the radial Alfv\'en crossing timescale of the remnant\footnote{In detail, the timescale to establish solid body can slightly exceed $t_{\rm A}$ (e.g., \citealt{Charbonneau&MacGregor92}).} given its large-scale magnetic field $B$,
\be
t_{\rm A} = \frac{R_{\rm ns}}{v_{\rm A}} \approx 1.0\left(\frac{B}{10^{14}\,\rm G}\right)^{-1}{\rm s}
\label{eq:tA}
\ee
where $v_{\rm A} = B/\sqrt{4\pi \rho}$ is the Alfv\'en velocity and $\rho \approx 3M_{\rm ns}/(4\pi R_{\rm ns}^{3})$ is the mean density of the NS.   Interestingly, the value of $t_{\rm A}$ for values of $B \sim 1-3\times 10^{14}$ G we find are needed to explain the blue KN is similar to the HMNS collapse time $t_{\rm rem} \approx 0.1-1$ s inferred independently from the KN emission (and also possibly the arrival delay of the gamma-ray burst).

The hyper-massive magnetar scenario makes predictions for how the blue KN emission will correlate with the GW-inferred  properties of the NS-NS binary progenitor and, potentially, the timescale of the gamma-ray emission.  In general, more massive binaries will create less-stable HMNS remnants that collapse to black holes earlier (smaller $t_{\rm rem}$), which (for fixed $B$) would reduce the quantity of the blue KN ejecta from the magnetar wind ($M_{\rm blue} \propto t_{\rm rem}$).  On the other hand, if $t_{\rm A}$ (eq.~\ref{eq:tA}) also plays a role in the collapse time, then a merger remnant of total mass otherwise similar to that in GW170817, but with a stronger magnetic field $B$, will produce both a higher KN ejecta velocity $v_{\rm B} \propto B^{2/3}$ and a shorter lifetime $t_{\rm rem} \propto B^{-1} \propto v_{\rm B}^{-3/2}$; the latter may correlate also with the timescale of the observed gamma-ray burst, assuming it is powered by an ultra-relativistic jet which is generated soon after black hole formation.

Although disk outflows provide a satisfactory explanation for the slower, red KN ejecta (e.g.~\citealt{Siegel&Metzger17}), it is important to address whether the latter also be explained by a magnetar wind with time-evolving properties.  If the magnetic field of the HMNS remnant were to grow in time, then Fig.~\ref{fig:wind} shows that $Y_{e}$ will decrease and $\dot{M}$ will increase  (for fixed $\Omega$), although the decaying neutrino luminosity as the remnant cools will counter the latter.  If  $Y_{e}$ were to decrease in time, this would be consistent with the observed late emergence of red/purple KN ejecta ($Y_{e} \lesssim 0.25$) following the earlier blue KN ejecta ($Y_{e} \gtrsim 0.25$).  However, an increasing magnetic field would also cause the wind velocity to grow in time (Fig.~\ref{fig:windevo}), which, as discussed in $\S\ref{sec:earlytime}$, would give rise to powerful internal shocks.  This interaction would cause a redistribution of the radial velocity stratification and might lead to mixing of the high- and low-$Y_{e}$ material (see below).  We conclude that a time-evolving magnetar wind could in principle supply both the blue and red KN ejecta, though detailed numerical modeling of the magnetized outflow including neutrino transport from the remnant will be needed to confirm this idea quantitatively.

We have discussed ways to observationally disentangle different sources for the ejecta based on the first few hours of the KN (Fig.~\ref{fig:KN}).  In particular, the KN luminosity will be substantially higher at these early times if the ejecta is re-heated above its launching point by internal shocks.  The latter is generally expected for any long-lived ejecta source, e.g.~a magnetar of lifetime $t_{\rm rem} \gtrsim 0.1-1$ s (e.g.~\citealt{Metzger+08b}) or an accretion disk of lifetime $\sim 1$ s (e.g.~\citealt{Fernandez&Metzger13}), and is not a unique prediction of a (failed or successful) ultra-relativistic jet \citep{Gottlieb+18}.  This will unfortunately make the source of early ejecta heating (jet, magnetar wind, disk outflow) challenging to identify.  Early-time blue emission from shocks is also degenerate with heating from free neutron decay in the outer ejecta layers \citep{Metzger+15}.  It is therefore important to explore any additional signatures of late-time ejecta heating or shocks.  

Several groups have argued that the entire KN emission from GW170817 can be explained by a single homogeneous ejecta with a low lanthanide mass fraction, e.g. $X_{\rm La} \approx 10^{-3}$ (e.g.~\citealt{Tanaka+17,Smartt+17,Waxman+17}).  However, as already discussed, endowing the ejecta with a low but non-zero lanthanide abundance appears to require fine tuning of the $Y_{e}$ distribution to be tightly peaked around $Y_{e} \approx 0.25$ (see also \citealt{Tanaka+17}).  

One loophole to creating a homogeneous low-$X_{\rm La}$ composition would be to mix separate ejecta components with vastly different lanthanide abundances, e.g.~combine 1\% of the mass with $Y_{e} \lesssim 0.25$ and $X_{\rm La} \approx 0.1$ with a much larger fraction $\gtrsim 99\%$ that was lanthanide-free ($Y_{e} \gtrsim 0.25$).   Importantly, however, {\it such mixing would need to occur after the $r$-process is complete, i.e. after free neutrons are captured into nuclei on a timescale of $t_{\rm capture} \sim 1$ s after ejection}.  While such late-time mixing would not be consistent with a dynamical origin for the ejecta, it could result naturally if the engine lifetime was indeed $t_{\rm rem} \gtrsim 0.1-1$ s $\sim t_{\rm capture}$, as predicted for a long-lived magnetar.  As already discussed, the moderate rise in the magnetar outflow velocity, even over its short lifetime, will indeed produce internal shocks within the ejecta \citep{Metzger+08b} that would also enable mixing.  Moderate variations in $Y_{e}$ are also expected with polar latitude in the wind \citep{Vlasov+14} as well as in time, both secularly as the HMNS cools and on shorter timescales due to periodic magnetic reconnection events instigated by opening of closed magnetic field lines by differential rotation and neutrino heating \citep{Thompson03}.


\acknowledgements
BDM is supported by NASA through the ATP program, grant numbers NNX16AB30G and NNX17AK43G.  EQ was supported in part by a Simons Investigator Award from the Simons Foundation and the Gordon and Betty Moore Foundation through Grant GBMF5076.  We thank Niccolo Bucciantini for previous collaborations and many useful discussions on this topic.

\section*{Appendix: Layered Kilonova Model}

Multi-velocity semi-analytic one-dimensional kilonova models have been considered in several previous works (e.g.~\citealt{Piran+13,Metzger17,Waxman+17}).  Following the merger and any subsequent shock heating, e.g. by the GRB jet, the radial velocity structure of the ejecta ultimately approaches homologous expansion (e.g.~\citealt{Rosswog+14}).  We approximate the distribution of mass with velocity greater than a value $v$ as a power-law of index $\beta$,
\be
M_{v} = M(v/v_{0})^{-\beta}, v \gtrsim v_0
\ee
where $M$ is the total mass, $v_{0}$ is the minimum ($\sim$ average) velocity.  
The thermal energy of each mass shell evolves as a function of time $t$ after the merger according to
\be
\frac{dE_{v}}{dt} = -\frac{E_{v}}{t} - \frac{E_{v}}{t_{d,v} + t_{ lc,v}} + \dot{Q}_{v},
\label{eq:dEdt}
\ee
where the first term accounts for PdV losses, the second term is the radiative luminosity, where 
\be
t_{ d} = \frac{3M_{v}\kappa}{4\pi c v t}
\ee
is the photon diffusion time through layer $v$ and $t_{\rm lc,v} = (v/c)t$ is the light crossing time.  We have modeled the opacity produced by $r$-process lines as a grey opacity of value $\kappa$.  \citet{Tanaka+17} show that bolometric luminosities are reasonably well-produced for $\kappa = 0.5$ cm$^{2}$ g$^{-1}$ for light $r$-process nuclei (lanthanide-free, $Y_{e} \gtrsim 0.25$) and $\kappa = 10$ cm$^{2}$ g$^{-1}$ (full lanthanide-fraction, $Y_{e} \lesssim 0.25$).

The final term in equation (\ref{eq:dEdt}) is the radioactive heating rate, which for $r$-process nuclei is reasonably approximated by \citep{Metzger+10,Roberts+11}
\be
\dot{Q}_{v} = \epsilon_{ th,v}M_{v}\dot{Q}_{1,v}(t/1{\rm d})^{-1.3},
\label{eq:qdotv}
\ee
where the normalization depends on both the thermalization efficiency $\epsilon_{\rm th,v}$ of the decay products and the overall normalization $\dot{Q}_{1,v}$ of the specific radioactive energy loss rate at $1$ day.  The thermalization efficiency varies with the density of the mass shell, with a high value at early times $\approx 0.5-0.7$ due to the efficient exchange of beta decay and gamma-ray products \citep{Metzger+10,Barnes+16,Hotokezaka+16,Waxman+17}.  The value of $\dot{Q}_{1,v}$ depends on the nucleosynthesis products and varies with $Y_{e}$ from $\approx 1.5\times 10^{10}$ ergs g$^{-1}$ s$^{-1}$ at low $Y_{e} \lesssim 0.3$ to a value $\approx 1-4\times 10^{10}$ ergs g$^{-1}$ s$^{-1}$ at higher $Y_{e}$.    In what follows we take $\dot{Q}_{1,v} = 2\times 10^{10}$ ergs g$^{-1}$ s$^{-1}$.

In some cases we also include radioactive heating from free neutrons (decay timescale $\tau_{\rm n} \approx 900$ s), which exist in the outermost layers ($M_{\rm n} \approx 10^{-4}-10^{-3}M_{\odot}$) of the ejecta and producing a radioactive heating rate (in lieu of eq.~\ref{eq:qdotv}) of
\be
\dot{Q}_{\rm v} = 3.2\times 10^{14}(1-2Y_{e})\,{\rm ergs\,g^{-1}\,s^{-1}}\exp(-t/\tau_{\rm n}), M_{\rm v} \lesssim M_{\rm n},
\ee 
where $1-2Y_{e}$ is the neutron fraction; as typically the outer layers of the ejecta have $Y_{e} \ll 1$ this factor is usually $\approx 1$.

Boundary conditions on the thermal energy are derived by fixing $E_{v} = v^{2}/2$ for each shell until times $t = t_0$, after which time the energy is evolved according to equation (\ref{eq:dEdt}).


\begin{thebibliography}{}

\bibitem[{{Abbott} {et~al.}(2017{\natexlab{a}}){Abbott}, {Abbott}, {Abbott},
  {Acernese}, {Ackley}, {Adams}, {Adams}, {Addesso}, {Adhikari}, {Adya}, \&
  et~al.}]{LIGO+17FERMI}
{Abbott}, B.~P., {Abbott}, R., {Abbott}, T.~D., {et~al.} 2017{\natexlab{a}},
  \href{http://dx.doi.org/10.3847/2041-8213/aa920c}{\JournalTitle{\apjl}, 848,
  L13}

\bibitem[{{Abbott} {et~al.}(2017{\natexlab{b}}){Abbott}, {Abbott}, {Abbott},
  {Acernese}, {Ackley}, {Adams}, {Adams}, {Addesso}, {Adhikari}, {Adya}, \&
  et~al.}]{LIGO+17DISCOVERY}
---. 2017{\natexlab{b}},
  \href{http://dx.doi.org/10.1103/PhysRevLett.119.161101}{\JournalTitle{Physical
  Review Letters}, 119, 161101}

\bibitem[{{Abbott} {et~al.}(2017{\natexlab{c}}){Abbott}, {Abbott}, {Abbott},
  {Acernese}, {Ackley}, {Adams}, {Adams}, {Addesso}, {Adhikari}, {Adya}, \&
  et~al.}]{LIGO+17CAPSTONE}
---. 2017{\natexlab{c}},
  \href{http://dx.doi.org/10.3847/2041-8213/aa91c9}{\JournalTitle{\apjl}, 848,
  L12}

\bibitem[{{Arcavi} {et~al.}(2017){Arcavi}, {Hosseinzadeh}, {Howell}, {McCully},
  {Poznanski}, {Kasen}, {Barnes}, {Zaltzman}, {Vasylyev}, {Maoz}, \&
  {Valenti}}]{Arcavi+17}
{Arcavi}, I., {Hosseinzadeh}, G., {Howell}, D.~A., {et~al.} 2017,
  \href{http://dx.doi.org/10.1038/nature24291}{\JournalTitle{\nat}, 551, 64}

\bibitem[{{Barnes} \& {Kasen}(2013)}]{Barnes&Kasen13}
{Barnes}, J., \& {Kasen}, D. 2013,
  \href{http://dx.doi.org/10.1088/0004-637X/775/1/18}{\JournalTitle{Astrophys.
  J.}, 775, 18}

\bibitem[{{Barnes} {et~al.}(2016){Barnes}, {Kasen}, {Wu}, \&
  {Mart{\'{\i}}nez-Pinedo}}]{Barnes+16}
{Barnes}, J., {Kasen}, D., {Wu}, M.-R., \& {Mart{\'{\i}}nez-Pinedo}, G. 2016,
  \href{http://dx.doi.org/10.3847/0004-637X/829/2/110}{\JournalTitle{Astrophys.
  J.}, 829, 110}

\bibitem[{{Bauswein} {et~al.}(2013){Bauswein}, {Goriely}, \&
  {Janka}}]{Bauswein+13}
{Bauswein}, A., {Goriely}, S., \& {Janka}, H.-T. 2013,
  \href{http://dx.doi.org/10.1088/0004-637X/773/1/78}{\JournalTitle{Astrophys.
  J.}, 773, 78}

\bibitem[{{Bauswein} {et~al.}(2017){Bauswein}, {Just}, {Janka}, \&
  {Stergioulas}}]{Bauswein+17}
{Bauswein}, A., {Just}, O., {Janka}, H.-T., \& {Stergioulas}, N. 2017,
  \href{http://dx.doi.org/10.3847/2041-8213/aa9994}{\JournalTitle{\apjl}, 850,
  L34}

\bibitem[{{Beloborodov}(2014)}]{Beloborodov14}
{Beloborodov}, A.~M. 2014,
  \href{http://dx.doi.org/10.1093/mnras/stt2140}{\JournalTitle{\mnras}, 438,
  169}

\bibitem[{{Berger}(2014)}]{Berger14}
{Berger}, E. 2014,
  \href{http://dx.doi.org/10.1146/annurev-astro-081913-035926}{\JournalTitle{Annu.
  Rev. Astron. Astrophys.}, 52, 43}

\bibitem[{{Bovard} {et~al.}(2017){Bovard}, {Martin}, {Guercilena}, {Arcones},
  {Rezzolla}, \& {Korobkin}}]{Bovard+17}
{Bovard}, L., {Martin}, D., {Guercilena}, F., {et~al.} 2017,
  \JournalTitle{ArXiv e-prints},
  \href{http://arxiv.org/abs/1709.09630}{{\sffamily arXiv:1709.09630 [gr-qc]}}

\bibitem[{{Braithwaite}(2007)}]{Braithwaite07}
{Braithwaite}, J. 2007,
  \href{http://dx.doi.org/10.1051/0004-6361:20065903}{\JournalTitle{\aap}, 469,
  275}

\bibitem[{{Bromberg} {et~al.}(2017){Bromberg}, {Tchekhovskoy}, {Gottlieb},
  {Nakar}, \& {Piran}}]{Bromberg+17}
{Bromberg}, O., {Tchekhovskoy}, A., {Gottlieb}, O., {Nakar}, E., \& {Piran}, T.
  2017, \JournalTitle{ArXiv e-prints},
  \href{http://arxiv.org/abs/1710.05897}{{\sffamily arXiv:1710.05897
  [astro-ph.HE]}}

\bibitem[{{Bucciantini} {et~al.}(2012){Bucciantini}, {Metzger}, {Thompson}, \&
  {Quataert}}]{Bucciantini+12}
{Bucciantini}, N., {Metzger}, B.~D., {Thompson}, T.~A., \& {Quataert}, E. 2012,
  \href{http://dx.doi.org/10.1111/j.1365-2966.2011.19810.x}{\JournalTitle{Mon.
  Not. R. Astron. Soc.}, 419, 1537}

\bibitem[{{Bucciantini} {et~al.}(2007){Bucciantini}, {Quataert}, {Arons},
  {Metzger}, \& {Thompson}}]{Bucciantini+07}
{Bucciantini}, N., {Quataert}, E., {Arons}, J., {Metzger}, B.~D., \&
  {Thompson}, T.~A. 2007,
  \href{http://dx.doi.org/10.1111/j.1365-2966.2007.12164.x}{\JournalTitle{\mnras},
  380, 1541}

\bibitem[{{Bucciantini} {et~al.}(2008){Bucciantini}, {Quataert}, {Arons},
  {Metzger}, \& {Thompson}}]{Bucciantini+08}
---. 2008,
  \href{http://dx.doi.org/10.1111/j.1745-3933.2007.00403.x}{\JournalTitle{\mnras},
  383, L25}

\bibitem[{{Bucciantini} {et~al.}(2009){Bucciantini}, {Quataert}, {Metzger},
  {Thompson}, {Arons}, \& {Del Zanna}}]{Bucciantini+09}
{Bucciantini}, N., {Quataert}, E., {Metzger}, B.~D., {et~al.} 2009,
  \href{http://dx.doi.org/10.1111/j.1365-2966.2009.14940.x}{\JournalTitle{\mnras},
  396, 2038}

\bibitem[{{Bucciantini} {et~al.}(2006){Bucciantini}, {Thompson}, {Arons},
  {Quataert}, \& {Del Zanna}}]{Bucciantini+06}
{Bucciantini}, N., {Thompson}, T.~A., {Arons}, J., {Quataert}, E., \& {Del
  Zanna}, L. 2006,
  \href{http://dx.doi.org/10.1111/j.1365-2966.2006.10217.x}{\JournalTitle{\mnras},
  368, 1717}

\bibitem[{{Buckley} {et~al.}(2017)}]{Buckley+17}
{Buckley}, {et~al.} 2017, \JournalTitle{MNRAS Letters},
  \href{http://arxiv.org/abs/XXX}{{\sffamily arXiv:XXX [astro-ph.HE]}}

\bibitem[{{Chang} \& {Murray}(2018)}]{Chang&Murray18}
{Chang}, P., \& {Murray}, N. 2018,
  \href{http://dx.doi.org/10.1093/mnrasl/slx186}{\JournalTitle{\mnras}, 474,
  L12}

\bibitem[{{Charbonneau} \& {MacGregor}(1992)}]{Charbonneau&MacGregor92}
{Charbonneau}, P., \& {MacGregor}, K.~B. 1992,
  \href{http://dx.doi.org/10.1086/186545}{\JournalTitle{\apjl}, 397, L63}

\bibitem[{{Chornock} {et~al.}(2017)}]{Chornock+17}
{Chornock}, R., {et~al.} 2017,
  \href{http://dx.doi.org/10.3847/2041-8213/aa905c}{\JournalTitle{\apjl}, 848,
  L19}

\bibitem[{{Coulter} {et~al.}(2017)}]{Coulter+17}
{Coulter}, D.~A., {et~al.} 2017, \JournalTitle{ArXiv e-prints},
  \href{http://arxiv.org/abs/1710.05452}{{\sffamily arXiv:1710.05452
  [astro-ph.HE]}}

\bibitem[{{Covino} {et~al.}(2017)}]{Covino+17}
{Covino}, S., {et~al.} 2017,
  \href{http://dx.doi.org/10.1038/s41550-017-0285-z}{\JournalTitle{Nature
  Astronomy}, 1, 791}

\bibitem[{{Cowperthwaite} {et~al.}(2017)}]{Cowperthwaite+17}
{Cowperthwaite}, P., {et~al.} 2017,
  \href{http://dx.doi.org/10.1038/ncomms12898}{\JournalTitle{ApJL}},
  \href{http://arxiv.org/abs/XXX}{{\sffamily arXiv:XXX [astro-ph.HE]}}

\bibitem[{{Dessart} {et~al.}(2009){Dessart}, {Ott}, {Burrows}, {Rosswog}, \&
  {Livne}}]{Dessart+09}
{Dessart}, L., {Ott}, C.~D., {Burrows}, A., {Rosswog}, S., \& {Livne}, E. 2009,
  \href{http://dx.doi.org/10.1088/0004-637X/690/2/1681}{\JournalTitle{Astrophys.
  J.}, 690, 1681}

\bibitem[{{D{\'{\i}}az} {et~al.}(2017)}]{Diaz+17}
{D{\'{\i}}az}, M.~C., {et~al.} 2017,
  \href{http://dx.doi.org/10.3847/2041-8213/aa9060}{\JournalTitle{\apjl}, 848,
  L29}

\bibitem[{{Dietrich} \& {Ujevic}(2017)}]{Dietrich&Ujevic17}
{Dietrich}, T., \& {Ujevic}, M. 2017,
  \href{http://dx.doi.org/10.1088/1361-6382/aa6bb0}{\JournalTitle{Classical and
  Quantum Gravity}, 34, 105014}

\bibitem[{{Dietrich} {et~al.}(2017){Dietrich}, {Ujevic}, {Tichy}, {Bernuzzi},
  \& {Br{\"u}gmann}}]{Dietrich+17}
{Dietrich}, T., {Ujevic}, M., {Tichy}, W., {Bernuzzi}, S., \& {Br{\"u}gmann},
  B. 2017,
  \href{http://dx.doi.org/10.1103/PhysRevD.95.024029}{\JournalTitle{\prd}, 95,
  024029}

\bibitem[{{Drout} {et~al.}(2017)}]{Drout+17}
{Drout}, M.~R., {et~al.} 2017, \JournalTitle{ArXiv e-prints},
  \href{http://arxiv.org/abs/1710.05443}{{\sffamily arXiv:1710.05443
  [astro-ph.HE]}}

\bibitem[{{Duffell} {et~al.}(2015){Duffell}, {Quataert}, \&
  {MacFadyen}}]{Duffell+15}
{Duffell}, P.~C., {Quataert}, E., \& {MacFadyen}, A.~I. 2015,
  \href{http://dx.doi.org/10.1088/0004-637X/813/1/64}{\JournalTitle{\apj}, 813,
  64}

\bibitem[{{Duncan} {et~al.}(1986){Duncan}, {Shapiro}, \&
  {Wasserman}}]{Duncan+86}
{Duncan}, R.~C., {Shapiro}, S.~L., \& {Wasserman}, I. 1986,
  \href{http://dx.doi.org/10.1086/164587}{\JournalTitle{Astrophys. J.}, 309,
  141}

\bibitem[{{Evans} {et~al.}(2017)}]{Evans+17}
{Evans}, P.~A., {et~al.} 2017, \JournalTitle{ArXiv e-prints},
  \href{http://arxiv.org/abs/1710.05437}{{\sffamily arXiv:1710.05437
  [astro-ph.HE]}}

\bibitem[{{Fern{\'a}ndez} \& {Metzger}(2013)}]{Fernandez&Metzger13}
{Fern{\'a}ndez}, R., \& {Metzger}, B.~D. 2013,
  \href{http://dx.doi.org/10.1093/mnras/stt1312}{\JournalTitle{Mon. Not. R.
  Astron. Soc.}, 435, 502}

\bibitem[{{Fern{\'a}ndez} \& {Metzger}(2016)}]{Fernandez&Metzger16}
---. 2016,
  \href{http://dx.doi.org/10.1146/annurev-nucl-102115-044819}{\JournalTitle{Annu.
  Rev. Nucl. Part. Sci.}, 66, 23}

\bibitem[{{Fong} {et~al.}(2015){Fong}, {Berger}, {Margutti}, \&
  {Zauderer}}]{Fong+15}
{Fong}, W., {Berger}, E., {Margutti}, R., \& {Zauderer}, B.~A. 2015,
  \href{http://dx.doi.org/10.1088/0004-637X/815/2/102}{\JournalTitle{Astrophys.
  J.}, 815, 102}

\bibitem[{{Fontes} {et~al.}(2017){Fontes}, {Fryer}, {Hungerford}, {Wollaeger},
  {Rosswog}, \& {Berger}}]{Fontes+17}
{Fontes}, C.~J., {Fryer}, C.~L., {Hungerford}, A.~L., {et~al.} 2017,
  \JournalTitle{ArXiv e-prints},
  \href{http://arxiv.org/abs/1702.02990}{{\sffamily arXiv:1702.02990
  [astro-ph.HE]}}

\bibitem[{{Gao} {et~al.}(2017){Gao}, {Cao}, {Ai}, \& {Zhang}}]{Gao+17}
{Gao}, H., {Cao}, Z., {Ai}, S., \& {Zhang}, B. 2017,
  \href{http://dx.doi.org/10.3847/2041-8213/aaa0c6}{\JournalTitle{\apjl}, 851,
  L45}

\bibitem[{{Goldstein} {et~al.}(2017)}]{Goldstein+17}
{Goldstein}, A., {et~al.} 2017,
  \href{http://dx.doi.org/10.3847/2041-8213/aa8f41}{\JournalTitle{\apjl}, 848,
  L14}

\bibitem[{{Goriely} {et~al.}(2015){Goriely}, {Bauswein}, {Just}, {Pllumbi}, \&
  {Janka}}]{Goriely+15}
{Goriely}, S., {Bauswein}, A., {Just}, O., {Pllumbi}, E., \& {Janka}, H.-T.
  2015, \href{http://dx.doi.org/10.1093/mnras/stv1526}{\JournalTitle{Mon. Not.
  R. Astron. Soc.}, 452, 3894}

\bibitem[{{Gottlieb} {et~al.}(2018){Gottlieb}, {Nakar}, \&
  {Piran}}]{Gottlieb+18}
{Gottlieb}, O., {Nakar}, E., \& {Piran}, T. 2018,
  \href{http://dx.doi.org/10.1093/mnras/stx2357}{\JournalTitle{\mnras}, 473,
  576}

\bibitem[{{Gottlieb} {et~al.}(2017){Gottlieb}, {Nakar}, {Piran}, \&
  {Hotokezaka}}]{Gottlieb+18b}
{Gottlieb}, O., {Nakar}, E., {Piran}, T., \& {Hotokezaka}, K. 2017,
  \JournalTitle{ArXiv e-prints},
  \href{http://arxiv.org/abs/1710.05896}{{\sffamily arXiv:1710.05896
  [astro-ph.HE]}}

\bibitem[{{Granot} {et~al.}(2017){Granot}, {Guetta}, \& {Gill}}]{Granot+17}
{Granot}, J., {Guetta}, D., \& {Gill}, R. 2017,
  \href{http://dx.doi.org/10.3847/2041-8213/aa991d}{\JournalTitle{\apjl}, 850,
  L24}

\bibitem[{{Grossman} {et~al.}(2014){Grossman}, {Korobkin}, {Rosswog}, \&
  {Piran}}]{Grossman+14}
{Grossman}, D., {Korobkin}, O., {Rosswog}, S., \& {Piran}, T. 2014,
  \href{http://dx.doi.org/10.1093/mnras/stt2503}{\JournalTitle{Mon. Not. R.
  Astron. Soc.}, 439, 757}

\bibitem[{{Hanauske} {et~al.}(2017){Hanauske}, {Takami}, {Bovard}, {Rezzolla},
  {Font}, {Galeazzi}, \& {St{\"o}cker}}]{Hanauske+17}
{Hanauske}, M., {Takami}, K., {Bovard}, L., {et~al.} 2017,
  \href{http://dx.doi.org/10.1103/PhysRevD.96.043004}{\JournalTitle{\prd}, 96,
  043004}

\bibitem[{{Hotokezaka} {et~al.}(2013){Hotokezaka}, {Kiuchi}, {Kyutoku},
  {Okawa}, {Sekiguchi}, {Shibata}, \& {Taniguchi}}]{Hotokezaka+13}
{Hotokezaka}, K., {Kiuchi}, K., {Kyutoku}, K., {et~al.} 2013,
  \href{http://dx.doi.org/10.1103/PhysRevD.87.024001}{\JournalTitle{Phys. Rev.
  D}, 87, 024001}

\bibitem[{{Hotokezaka} {et~al.}(2016){Hotokezaka}, {Wanajo}, {Tanaka}, {Bamba},
  {Terada}, \& {Piran}}]{Hotokezaka+16}
{Hotokezaka}, K., {Wanajo}, S., {Tanaka}, M., {et~al.} 2016,
  \href{http://dx.doi.org/10.1093/mnras/stw404}{\JournalTitle{\mnras}, 459, 35}

\bibitem[{{Hu} {et~al.}(2017)}]{Hu+17}
{Hu}, L., {et~al.} 2017, \JournalTitle{ArXiv e-prints},
  \href{http://arxiv.org/abs/1710.05462}{{\sffamily arXiv:1710.05462
  [astro-ph.HE]}}

\bibitem[{{Just} {et~al.}(2015){Just}, {Bauswein}, {Pulpillo}, {Goriely}, \&
  {Janka}}]{Just+15}
{Just}, O., {Bauswein}, A., {Pulpillo}, R.~A., {Goriely}, S., \& {Janka}, H.-T.
  2015, \href{http://dx.doi.org/10.1093/mnras/stv009}{\JournalTitle{Mon. Not.
  R. Astron. Soc.}, 448, 541}

\bibitem[{{Kaplan} {et~al.}(2014){Kaplan}, {Ott}, {O'Connor}, {Kiuchi},
  {Roberts}, \& {Duez}}]{Kaplan+14}
{Kaplan}, J.~D., {Ott}, C.~D., {O'Connor}, E.~P., {et~al.} 2014,
  \href{http://dx.doi.org/10.1088/0004-637X/790/1/19}{\JournalTitle{Astrophys.
  J.}, 790, 19}

\bibitem[{{Kasen} {et~al.}(2013){Kasen}, {Badnell}, \& {Barnes}}]{Kasen+13}
{Kasen}, D., {Badnell}, N.~R., \& {Barnes}, J. 2013, \JournalTitle{ApJ,
  submitted, arXiv:1303.5788}, \href{http://arxiv.org/abs/1303.5788}{{\sffamily
  arXiv:1303.5788 [astro-ph.HE]}}

\bibitem[{{Kasen} {et~al.}(2017){Kasen}, {Metzger}, {Barnes}, {Ramirez-Ruiz},
  \& {Quataert}}]{Kasen+17}
{Kasen}, D., {Metzger}, B., {Barnes}, J., {Ramirez-Ruiz}, \& {Quataert}, E.
  2017, \href{http://dx.doi.org/10.1038/ncomms12898}{\JournalTitle{Nature}},
  \href{http://arxiv.org/abs/XXX}{{\sffamily arXiv:XXX [astro-ph.HE]}}

\bibitem[{{Kasliwal} \& {Nissanke}(2014)}]{Kasliwal&Nissanke14}
{Kasliwal}, M.~M., \& {Nissanke}, S. 2014,
  \href{http://dx.doi.org/10.1088/2041-8205/789/1/L5}{\JournalTitle{\apjl},
  789, L5}

\bibitem[{{Kasliwal} {et~al.}(2017)}]{Kasliwal+17}
{Kasliwal}, M.~M., {et~al.} 2017, \JournalTitle{ArXiv e-prints},
  \href{http://arxiv.org/abs/1710.05436}{{\sffamily arXiv:1710.05436
  [astro-ph.HE]}}

\bibitem[{{Kilpatrick} {et~al.}(2017)}]{Kilpatrick+17}
{Kilpatrick}, C.~D., {et~al.} 2017, \JournalTitle{ArXiv e-prints},
  \href{http://arxiv.org/abs/1710.05434}{{\sffamily arXiv:1710.05434
  [astro-ph.HE]}}

\bibitem[{{Kiuchi} {et~al.}(2015){Kiuchi}, {Sekiguchi}, {Kyutoku}, {Shibata},
  {Taniguchi}, \& {Wada}}]{Kiuchi+15}
{Kiuchi}, K., {Sekiguchi}, Y., {Kyutoku}, K., {et~al.} 2015,
  \href{http://dx.doi.org/10.1103/PhysRevD.92.064034}{\JournalTitle{Phys. Rev.
  D}, 92, 064034}

\bibitem[{{Komissarov} \& {Barkov}(2007)}]{Komissarov&Barkov07}
{Komissarov}, S.~S., \& {Barkov}, M.~V. 2007,
  \href{http://dx.doi.org/10.1111/j.1365-2966.2007.12485.x}{\JournalTitle{\mnras},
  382, 1029}

\bibitem[{{Kulkarni}(2005)}]{Kulkarni05}
{Kulkarni}, S.~R. 2005, \JournalTitle{ArXiv e-prints},
  \href{http://arxiv.org/abs/arXiv:astro-ph/0510256}{{\sffamily
  arXiv:astro-ph/0510256}}

\bibitem[{{Li} \& {Paczy{\'n}ski}(1998)}]{Li&Paczynski98}
{Li}, L.-X., \& {Paczy{\'n}ski}, B. 1998,
  \href{http://dx.doi.org/10.1086/311680}{\JournalTitle{Astrophys. J. Lett.},
  507, L59}

\bibitem[{{Lippuner} {et~al.}(2017){Lippuner}, {Fern{\'a}ndez}, {Roberts},
  {Foucart}, {Kasen}, {Metzger}, \& {Ott}}]{Lippuner+17}
{Lippuner}, J., {Fern{\'a}ndez}, R., {Roberts}, L.~F., {et~al.} 2017,
  \href{http://dx.doi.org/10.1093/mnras/stx1987}{\JournalTitle{\mnras}, 472,
  904}

\bibitem[{{Lippuner} \& {Roberts}(2015)}]{Lippuner&Roberts15}
{Lippuner}, J., \& {Roberts}, L.~F. 2015,
  \href{http://dx.doi.org/10.1088/0004-637X/815/2/82}{\JournalTitle{Astrophys.
  J.}, 815, 82}

\bibitem[{{Ma} {et~al.}(2017){Ma}, {Jiang}, {Wang}, {Jin}, {Fan}, \&
  {Wei}}]{Ma+17}
{Ma}, P.-X., {Jiang}, J.-L., {Wang}, H., {et~al.} 2017, \JournalTitle{ArXiv
  e-prints}, \href{http://arxiv.org/abs/1711.05565}{{\sffamily arXiv:1711.05565
  [astro-ph.HE]}}

\bibitem[{{Margalit} \& {Metzger}(2017)}]{Margalit&Metzger17}
{Margalit}, B., \& {Metzger}, B.~D. 2017,
  \href{http://dx.doi.org/10.3847/2041-8213/aa991c}{\JournalTitle{\apjl}, 850,
  L19}

\bibitem[{{Margutti} {et~al.}(2017)}]{Margutti+17}
{Margutti}, M., {et~al.} 2017,
  \href{http://dx.doi.org/10.1038/ncomms12898}{\JournalTitle{ApJL accepted}},
  \href{http://arxiv.org/abs/XXX}{{\sffamily arXiv:XXX [astro-ph.HE]}}

\bibitem[{{Martin} {et~al.}(2015){Martin}, {Perego}, {Arcones}, {Thielemann},
  {Korobkin}, \& {Rosswog}}]{Martin+15}
{Martin}, D., {Perego}, A., {Arcones}, A., {et~al.} 2015,
  \href{http://dx.doi.org/10.1088/0004-637X/813/1/2}{\JournalTitle{Astrophys.
  J.}, 813, 2}

\bibitem[{{Mart{\'{\i}}nez-Pinedo} {et~al.}(2012){Mart{\'{\i}}nez-Pinedo},
  {Fischer}, {Lohs}, \& {Huther}}]{MartinezPinedo+12}
{Mart{\'{\i}}nez-Pinedo}, G., {Fischer}, T., {Lohs}, A., \& {Huther}, L. 2012,
  \href{http://dx.doi.org/10.1103/PhysRevLett.109.251104}{\JournalTitle{Phys.
  Rev. Lett.}, 109, 251104}

\bibitem[{{McCully} {et~al.}(2017)}]{McCully+17}
{McCully}, C., {et~al.} 2017,
  \href{http://dx.doi.org/10.3847/2041-8213/aa9111}{\JournalTitle{\apjl}, 848,
  L32}

\bibitem[{{Metzger}(2017)}]{Metzger17}
{Metzger}, B.~D. 2017,
  \href{http://dx.doi.org/10.1007/s41114-017-0006-z}{\JournalTitle{Living
  Reviews in Relativity}, 20, 3}

\bibitem[{{Metzger} {et~al.}(2015){Metzger}, {Bauswein}, {Goriely}, \&
  {Kasen}}]{Metzger+15}
{Metzger}, B.~D., {Bauswein}, A., {Goriely}, S., \& {Kasen}, D. 2015,
  \href{http://dx.doi.org/10.1093/mnras/stu2225}{\JournalTitle{Mon. Not. R.
  Astron. Soc.}, 446, 1115}

\bibitem[{{Metzger} \& {Fern{\'a}ndez}(2014)}]{Metzger&Fernandez14}
{Metzger}, B.~D., \& {Fern{\'a}ndez}, R. 2014,
  \href{http://dx.doi.org/10.1093/mnras/stu802}{\JournalTitle{Mon. Not. R.
  Astron. Soc.}, 441, 3444}

\bibitem[{{Metzger} {et~al.}(2011){Metzger}, {Giannios}, {Thompson},
  {Bucciantini}, \& {Quataert}}]{Metzger+11}
{Metzger}, B.~D., {Giannios}, D., {Thompson}, T.~A., {Bucciantini}, N., \&
  {Quataert}, E. 2011,
  \href{http://dx.doi.org/10.1111/j.1365-2966.2011.18280.x}{\JournalTitle{Mon.
  Not. R. Astron. Soc.}, 413, 2031}

\bibitem[{{Metzger} {et~al.}(2008{\natexlab{a}}){Metzger}, {Piro}, \&
  {Quataert}}]{Metzger+08c}
{Metzger}, B.~D., {Piro}, A.~L., \& {Quataert}, E. 2008{\natexlab{a}},
  \href{http://dx.doi.org/10.1111/j.1365-2966.2008.13789.x}{\JournalTitle{\mnras},
  390, 781}

\bibitem[{{Metzger} {et~al.}(2009){Metzger}, {Piro}, \&
  {Quataert}}]{Metzger+09}
---. 2009,
  \href{http://dx.doi.org/10.1111/j.1365-2966.2008.14380.x}{\JournalTitle{Mon.
  Not. R. Astron. Soc.}, 396, 304}

\bibitem[{{Metzger} {et~al.}(2008{\natexlab{b}}){Metzger}, {Quataert}, \&
  {Thompson}}]{Metzger+08b}
{Metzger}, B.~D., {Quataert}, E., \& {Thompson}, T.~A. 2008{\natexlab{b}},
  \href{http://dx.doi.org/10.1111/j.1365-2966.2008.12923.x}{\JournalTitle{Mon.
  Not. R. Astron. Soc.}, 385, 1455}

\bibitem[{{Metzger} {et~al.}(2007){Metzger}, {Thompson}, \&
  {Quataert}}]{Metzger+07}
{Metzger}, B.~D., {Thompson}, T.~A., \& {Quataert}, E. 2007,
  \href{http://dx.doi.org/10.1086/512059}{\JournalTitle{Astrophys. J.}, 659,
  561}

\bibitem[{{Metzger} {et~al.}(2008{\natexlab{c}}){Metzger}, {Thompson}, \&
  {Quataert}}]{Metzger+08a}
---. 2008{\natexlab{c}},
  \href{http://dx.doi.org/10.1086/526418}{\JournalTitle{Astrophys. J.}, 676,
  1130}

\bibitem[{{Metzger} {et~al.}(2010){Metzger}, {Mart{\'{\i}}nez-Pinedo},
  {Darbha}, {Quataert}, {Arcones}, {Kasen}, {Thomas}, {Nugent}, {Panov}, \&
  {Zinner}}]{Metzger+10}
{Metzger}, B.~D., {Mart{\'{\i}}nez-Pinedo}, G., {Darbha}, S., {et~al.} 2010,
  \href{http://dx.doi.org/10.1111/j.1365-2966.2010.16864.x}{\JournalTitle{Mon.
  Not. R. Astron. Soc.}, 406, 2650}

\bibitem[{{Michel}(1969)}]{Michel69}
{Michel}, F.~C. 1969,
  \href{http://dx.doi.org/10.1086/150233}{\JournalTitle{\apj}, 158, 727}

\bibitem[{{Murguia-Berthier} {et~al.}(2016){Murguia-Berthier}, {Ramirez-Ruiz},
  {Montes}, {De Colle}, {Rezzolla}, {Rosswog}, {Takami}, {Perego}, \&
  {Lee}}]{Murguia-Berthier+16}
{Murguia-Berthier}, A., {Ramirez-Ruiz}, E., {Montes}, G., {et~al.} 2016,
  \href{http://dx.doi.org/10.3847/2041-8213/aa5b9e}{\JournalTitle{Astrophys. J.
  Lett.}, 835, L34}

\bibitem[{{Murguia-Berthier} {et~al.}(2017){Murguia-Berthier}, {Ramirez-Ruiz},
  {Kilpatrick}, {Foley}, {Kasen}, {Lee}, {Piro}, {Coulter}, {Drout}, {Madore},
  {Shappee}, {Pan}, {Prochaska}, {Rest}, {Rojas-Bravo}, {Siebert}, \&
  {Simon}}]{Murguia-Berthier+17}
{Murguia-Berthier}, A., {Ramirez-Ruiz}, E., {Kilpatrick}, C.~D., {et~al.} 2017,
  \href{http://dx.doi.org/10.3847/2041-8213/aa91b3}{\JournalTitle{\apjl}, 848,
  L34}

\bibitem[{{Nakar}(2007)}]{Nakar07}
{Nakar}, E. 2007,
  \href{http://dx.doi.org/10.1016/j.physrep.2007.02.005}{\JournalTitle{Phys.
  Rep.}, 442, 166}

\bibitem[{{Nakar} \& {Sari}(2012)}]{Nakar&Sari12}
{Nakar}, E., \& {Sari}, R. 2012,
  \href{http://dx.doi.org/10.1088/0004-637X/747/2/88}{\JournalTitle{\apj}, 747,
  88}

\bibitem[{{Nicholl} {et~al.}(2017)}]{Nicholl+17}
{Nicholl}, M., {et~al.} 2017,
  \href{http://dx.doi.org/10.3847/2041-8213/aa9029}{\JournalTitle{\apjl}, 848,
  L18}

\bibitem[{{Oechslin} \& {Janka}(2006)}]{Oechslin&Janka06}
{Oechslin}, R., \& {Janka}, H.-T. 2006,
  \href{http://dx.doi.org/10.1111/j.1365-2966.2006.10238.x}{\JournalTitle{Mon.
  Not. R. Astron. Soc.}, 368, 1489}

\bibitem[{{Parfrey} {et~al.}(2016){Parfrey}, {Spitkovsky}, \&
  {Beloborodov}}]{Parfrey+16}
{Parfrey}, K., {Spitkovsky}, A., \& {Beloborodov}, A.~M. 2016,
  \href{http://dx.doi.org/10.3847/0004-637X/822/1/33}{\JournalTitle{\apj}, 822,
  33}

\bibitem[{{Perego} {et~al.}(2017){Perego}, {Radice}, \& {Bernuzzi}}]{Perego+17}
{Perego}, A., {Radice}, D., \& {Bernuzzi}, S. 2017,
  \href{http://dx.doi.org/10.3847/2041-8213/aa9ab9}{\JournalTitle{\apjl}, 850,
  L37}

\bibitem[{{Perego} {et~al.}(2014){Perego}, {Rosswog}, {Cabez{\'o}n},
  {Korobkin}, {K{\"a}ppeli}, {Arcones}, \& {Liebend{\"o}rfer}}]{Perego+14}
{Perego}, A., {Rosswog}, S., {Cabez{\'o}n}, R.~M., {et~al.} 2014,
  \href{http://dx.doi.org/10.1093/mnras/stu1352}{\JournalTitle{Mon. Not. R.
  Astron. Soc.}, 443, 3134}

\bibitem[{{Pian} {et~al.}(2017)}]{Pian+17}
{Pian}, E., {et~al.} 2017, \JournalTitle{Science},
  \href{http://arxiv.org/abs/XXX}{{\sffamily arXiv:XXX [astro-ph.HE]}}

\bibitem[{{Piran} {et~al.}(2013){Piran}, {Nakar}, \& {Rosswog}}]{Piran+13}
{Piran}, T., {Nakar}, E., \& {Rosswog}, S. 2013,
  \href{http://dx.doi.org/10.1093/mnras/stt037}{\JournalTitle{Mon. Not. R.
  Astron. Soc.}, 430, 2121}

\bibitem[{{Piro} \& {Kollmeier}(2017)}]{Piro&Kollmeier17}
{Piro}, A.~L., \& {Kollmeier}, J.~A. 2017, \JournalTitle{ArXiv e-prints},
  \href{http://arxiv.org/abs/1710.05822}{{\sffamily arXiv:1710.05822
  [astro-ph.HE]}}

\bibitem[{{Pons} {et~al.}(1999){Pons}, {Reddy}, {Prakash}, {Lattimer}, \&
  {Miralles}}]{Pons+99}
{Pons}, J.~A., {Reddy}, S., {Prakash}, M., {Lattimer}, J.~M., \& {Miralles},
  J.~A. 1999, \href{http://dx.doi.org/10.1086/306889}{\JournalTitle{\apj}, 513,
  780}

\bibitem[{{Pooley} {et~al.}(2017){Pooley}, {Kumar}, \& {Wheeler}}]{Pooley+17}
{Pooley}, D., {Kumar}, P., \& {Wheeler}, J.~C. 2017, \JournalTitle{ArXiv
  e-prints}, \href{http://arxiv.org/abs/1712.03240}{{\sffamily arXiv:1712.03240
  [astro-ph.HE]}}

\bibitem[{{Price} \& {Rosswog}(2006)}]{Price&Rosswog06}
{Price}, D.~J., \& {Rosswog}, S. 2006,
  \href{http://dx.doi.org/10.1126/science.1125201}{\JournalTitle{Science}, 312,
  719}

\bibitem[{{Qian} \& {Woosley}(1996)}]{Qian&Woosley96}
{Qian}, Y., \& {Woosley}, S.~E. 1996,
  \href{http://dx.doi.org/10.1086/177973}{\JournalTitle{Astrophys. J.}, 471,
  331}

\bibitem[{{Qian} {et~al.}(1993){Qian}, {Fuller}, {Mathews}, {Mayle}, {Wilson},
  \& {Woosley}}]{Qian+93}
{Qian}, Y.-Z., {Fuller}, G.~M., {Mathews}, G.~J., {et~al.} 1993,
  \href{http://dx.doi.org/10.1103/PhysRevLett.71.1965}{\JournalTitle{Physical
  Review Letters}, 71, 1965}

\bibitem[{{Radice}(2017)}]{Radice17}
{Radice}, D. 2017,
  \href{http://dx.doi.org/10.3847/2041-8213/aa6483}{\JournalTitle{\apjl}, 838,
  L2}

\bibitem[{{Radice} {et~al.}(2016){Radice}, {Galeazzi}, {Lippuner}, {Roberts},
  {Ott}, \& {Rezzolla}}]{Radice+16}
{Radice}, D., {Galeazzi}, F., {Lippuner}, J., {et~al.} 2016,
  \href{http://dx.doi.org/10.1093/mnras/stw1227}{\JournalTitle{\mnras}, 460,
  3255}

\bibitem[{{Radice} {et~al.}(2017){Radice}, {Perego}, {Zappa}, \&
  {Bernuzzi}}]{Radice+17}
{Radice}, D., {Perego}, A., {Zappa}, F., \& {Bernuzzi}, S. 2017,
  \JournalTitle{ArXiv e-prints},
  \href{http://arxiv.org/abs/1711.03647}{{\sffamily arXiv:1711.03647
  [astro-ph.HE]}}

\bibitem[{{Rezzolla} {et~al.}(2017){Rezzolla}, {Most}, \& {Weih}}]{Rezzolla+17}
{Rezzolla}, L., {Most}, E.~R., \& {Weih}, L.~R. 2017, \JournalTitle{ArXiv
  e-prints}, \href{http://arxiv.org/abs/1711.00314}{{\sffamily arXiv:1711.00314
  [astro-ph.HE]}}

\bibitem[{{Roberts} {et~al.}(2011){Roberts}, {Kasen}, {Lee}, \&
  {Ramirez-Ruiz}}]{Roberts+11}
{Roberts}, L.~F., {Kasen}, D., {Lee}, W.~H., \& {Ramirez-Ruiz}, E. 2011,
  \href{http://dx.doi.org/10.1088/2041-8205/736/1/L21}{\JournalTitle{Astrophys.
  J. Lett.}, 736, L21}

\bibitem[{{Roberts} {et~al.}(2012){Roberts}, {Reddy}, \& {Shen}}]{Roberts+12}
{Roberts}, L.~F., {Reddy}, S., \& {Shen}, G. 2012,
  \href{http://dx.doi.org/10.1103/PhysRevC.86.065803}{\JournalTitle{Phys. Rev.
  C}, 86, 065803}

\bibitem[{{Rosswog} {et~al.}(2014){Rosswog}, {Korobkin}, {Arcones},
  {Thielemann}, \& {Piran}}]{Rosswog+14}
{Rosswog}, S., {Korobkin}, O., {Arcones}, A., {Thielemann}, F.-K., \& {Piran},
  T. 2014, \href{http://dx.doi.org/10.1093/mnras/stt2502}{\JournalTitle{Mon.
  Not. R. Astron. Soc.}, 439, 744}

\bibitem[{{Rosswog} {et~al.}(1999){Rosswog}, {Liebend{\"o}rfer}, {Thielemann},
  {Davies}, {Benz}, \& {Piran}}]{Rosswog+99}
{Rosswog}, S., {Liebend{\"o}rfer}, M., {Thielemann}, F.-K., {et~al.} 1999,
  \JournalTitle{\aap}, 341, 499

\bibitem[{{Rosswog} {et~al.}(2017){Rosswog}, {Sollerman}, {Feindt}, {Goobar},
  {Korobkin}, {Fremling}, \& {Kasliwal}}]{Rosswog+17}
{Rosswog}, S., {Sollerman}, J., {Feindt}, U., {et~al.} 2017,
  \JournalTitle{ArXiv e-prints},
  \href{http://arxiv.org/abs/1710.05445}{{\sffamily arXiv:1710.05445
  [astro-ph.HE]}}

\bibitem[{{Ruiz} {et~al.}(2017){Ruiz}, {Shapiro}, \& {Tsokaros}}]{Ruiz+17}
{Ruiz}, M., {Shapiro}, S.~L., \& {Tsokaros}, A. 2017, \JournalTitle{ArXiv
  e-prints}, \href{http://arxiv.org/abs/1711.00473}{{\sffamily arXiv:1711.00473
  [astro-ph.HE]}}

\bibitem[{{Savchenko} {et~al.}(2017)}]{Savchenko+17}
{Savchenko}, V., {et~al.} 2017,
  \href{http://dx.doi.org/10.3847/2041-8213/aa8f94}{\JournalTitle{\apjl}, 848,
  L15}

\bibitem[{{Sekiguchi} {et~al.}(2016){Sekiguchi}, {Kiuchi}, {Kyutoku},
  {Shibata}, \& {Taniguchi}}]{Sekiguchi+16}
{Sekiguchi}, Y., {Kiuchi}, K., {Kyutoku}, K., {Shibata}, M., \& {Taniguchi}, K.
  2016,
  \href{http://dx.doi.org/10.1103/PhysRevD.93.124046}{\JournalTitle{\prd}, 93,
  124046}

\bibitem[{{Shappee} {et~al.}(2017)}]{Shappee+17}
{Shappee}, B.~J., {et~al.} 2017, \JournalTitle{ArXiv e-prints},
  \href{http://arxiv.org/abs/1710.05432}{{\sffamily arXiv:1710.05432
  [astro-ph.HE]}}

\bibitem[{{Shibata} {et~al.}(2017){Shibata}, {Fujibayashi}, {Hotokezaka},
  {Kiuchi}, {Kyutoku}, {Sekiguchi}, \& {Tanaka}}]{Shibata+17}
{Shibata}, M., {Fujibayashi}, S., {Hotokezaka}, K., {et~al.} 2017,
  \JournalTitle{ArXiv e-prints},
  \href{http://arxiv.org/abs/1710.07579}{{\sffamily arXiv:1710.07579
  [astro-ph.HE]}}

\bibitem[{{Shibata} \& {Kiuchi}(2017)}]{Shibata&Kiuchi17}
{Shibata}, M., \& {Kiuchi}, K. 2017,
  \href{http://dx.doi.org/10.1103/PhysRevD.95.123003}{\JournalTitle{\prd}, 95,
  123003}

\bibitem[{{Shibata} \& {Taniguchi}(2006)}]{Shibata&Taniguchi06}
{Shibata}, M., \& {Taniguchi}, K. 2006,
  \href{http://dx.doi.org/10.1103/PhysRevD.73.064027}{\JournalTitle{Phys. Rev.
  D}, 73, 064027}

\bibitem[{{Siegel} {et~al.}(2013){Siegel}, {Ciolfi}, {Harte}, \&
  {Rezzolla}}]{Siegel+13}
{Siegel}, D.~M., {Ciolfi}, R., {Harte}, A.~I., \& {Rezzolla}, L. 2013,
  \href{http://dx.doi.org/10.1103/PhysRevD.87.121302}{\JournalTitle{Phys. Rev.
  D}, 87, 121302}

\bibitem[{{Siegel} {et~al.}(2014){Siegel}, {Ciolfi}, \& {Rezzolla}}]{Siegel+14}
{Siegel}, D.~M., {Ciolfi}, R., \& {Rezzolla}, L. 2014,
  \href{http://dx.doi.org/10.1088/2041-8205/785/1/L6}{\JournalTitle{Astrophys.
  J. Lett.}, 785, L6}

\bibitem[{{Siegel} \& {Metzger}(2017{\natexlab{a}})}]{Siegel&Metzger17b}
{Siegel}, D.~M., \& {Metzger}, B.~D. 2017{\natexlab{a}}, \JournalTitle{ArXiv
  e-prints}, \href{http://arxiv.org/abs/1711.00868}{{\sffamily arXiv:1711.00868
  [astro-ph.HE]}}

\bibitem[{{Siegel} \& {Metzger}(2017{\natexlab{b}})}]{Siegel&Metzger17}
---. 2017{\natexlab{b}}, \JournalTitle{ArXiv e-prints},
  \href{http://arxiv.org/abs/1705.05473}{{\sffamily arXiv:1705.05473
  [astro-ph.HE]}}

\bibitem[{{Smartt} {et~al.}(2017)}]{Smartt+17}
{Smartt}, S., {et~al.} 2017, \JournalTitle{Nature},
  \href{http://arxiv.org/abs/XXX}{{\sffamily arXiv:XXX [astro-ph.HE]}}

\bibitem[{{Soares-Santos} {et~al.}(2017)}]{Soares-Santos+17}
{Soares-Santos}, M., {et~al.} 2017,
  \href{http://dx.doi.org/10.3847/2041-8213/aa9059}{\JournalTitle{\apjl}, 848,
  L16}

\bibitem[{{Tanaka} \& {Hotokezaka}(2013)}]{Tanaka&Hotokezaka13}
{Tanaka}, M., \& {Hotokezaka}, K. 2013,
  \href{http://dx.doi.org/10.1088/0004-637X/775/2/113}{\JournalTitle{Astrophys.
  J.}, 775, 113}

\bibitem[{{Tanaka} {et~al.}(2017{\natexlab{a}})}]{Tanaka+17b}
{Tanaka}, M., {et~al.} 2017{\natexlab{a}},
  \href{http://dx.doi.org/10.1093/pasj/psx121}{\JournalTitle{\pasj}, 69, 102}

\bibitem[{{Tanaka} {et~al.}(2017{\natexlab{b}}){Tanaka}, {Kato}, {Gaigalas},
  {Rynkun}, {Radziute}, {Wanajo}, {Sekiguchi}, {Nakamura}, {Tanuma},
  {Murakami}, \& {Sakaue}}]{Tanaka+17}
{Tanaka}, M., {Kato}, D., {Gaigalas}, G., {et~al.} 2017{\natexlab{b}},
  \JournalTitle{ArXiv e-prints},
  \href{http://arxiv.org/abs/1708.09101}{{\sffamily arXiv:1708.09101
  [astro-ph.HE]}}

\bibitem[{{Tanvir} {et~al.}(2017)}]{Tanvir+17}
{Tanvir}, N.~R., {et~al.} 2017,
  \href{http://dx.doi.org/10.3847/2041-8213/aa90b6}{\JournalTitle{\apjl}, 848,
  L27}

\bibitem[{{Thompson}(2003)}]{Thompson03}
{Thompson}, T.~A. 2003,
  \href{http://dx.doi.org/10.1086/374261}{\JournalTitle{Astrophys. J. Lett.},
  585, L33}

\bibitem[{{Thompson} {et~al.}(2001){Thompson}, {Burrows}, \&
  {Meyer}}]{Thompson+01}
{Thompson}, T.~A., {Burrows}, A., \& {Meyer}, B.~S. 2001,
  \href{http://dx.doi.org/10.1086/323861}{\JournalTitle{Astrophys. J.}, 562,
  887}

\bibitem[{{Thompson} {et~al.}(2004){Thompson}, {Chang}, \&
  {Quataert}}]{Thompson+04}
{Thompson}, T.~A., {Chang}, P., \& {Quataert}, E. 2004,
  \href{http://dx.doi.org/10.1086/421969}{\JournalTitle{Astrophys. J.}, 611,
  380}

\bibitem[{{Thompson} \& {ud-Doula}(2017)}]{Thompson&udDoula17}
{Thompson}, T.~A., \& {ud-Doula}, A. 2017, \JournalTitle{ArXiv e-prints},
  \href{http://arxiv.org/abs/1709.03997}{{\sffamily arXiv:1709.03997
  [astro-ph.HE]}}

\bibitem[{{Utsumi} {et~al.}(2017)}]{Utsumi+17}
{Utsumi}, Y., {et~al.} 2017,
  \href{http://dx.doi.org/10.1093/pasj/psx118}{\JournalTitle{\pasj}, 69, 101}

\bibitem[{{Villar} {et~al.}(2017){Villar}, {Guillochon}, {Berger}, {Metzger},
  {Cowperthwaite}, {Nicholl}, {Alexander}, {Blanchard}, {Chornock},
  {Eftekhari}, {Fong}, {Margutti}, \& {Williams}}]{Villar+17}
{Villar}, V.~A., {Guillochon}, J., {Berger}, E., {et~al.} 2017,
  \href{http://dx.doi.org/10.3847/2041-8213/aa9c84}{\JournalTitle{\apjl}, 851,
  L21}

\bibitem[{{Vlasov} {et~al.}(2017){Vlasov}, {Metzger}, {Lippuner}, {Roberts}, \&
  {Thompson}}]{Vlasov+17}
{Vlasov}, A.~D., {Metzger}, B.~D., {Lippuner}, J., {Roberts}, L.~F., \&
  {Thompson}, T.~A. 2017,
  \href{http://dx.doi.org/10.1093/mnras/stx478}{\JournalTitle{\mnras}, 468,
  1522}

\bibitem[{{Vlasov} {et~al.}(2014){Vlasov}, {Metzger}, \&
  {Thompson}}]{Vlasov+14}
{Vlasov}, A.~D., {Metzger}, B.~D., \& {Thompson}, T.~A. 2014,
  \href{http://dx.doi.org/10.1093/mnras/stu1667}{\JournalTitle{Mon. Not. R.
  Astron. Soc.}, 444, 3537}

\bibitem[{{Wanajo} {et~al.}(2014){Wanajo}, {Sekiguchi}, {Nishimura}, {Kiuchi},
  {Kyutoku}, \& {Shibata}}]{Wanajo+14}
{Wanajo}, S., {Sekiguchi}, Y., {Nishimura}, N., {et~al.} 2014,
  \href{http://dx.doi.org/10.1088/2041-8205/789/2/L39}{\JournalTitle{Astrophys.
  J. Lett.}, 789, L39}

\bibitem[{{Wanderman} \& {Piran}(2015)}]{Wanderman&Piran15}
{Wanderman}, D., \& {Piran}, T. 2015,
  \href{http://dx.doi.org/10.1093/mnras/stv123}{\JournalTitle{\mnras}, 448,
  3026}

\bibitem[{{Waxman} {et~al.}(2017){Waxman}, {Ofek}, {Kushnir}, \&
  {Gal-Yam}}]{Waxman+17}
{Waxman}, E., {Ofek}, E., {Kushnir}, D., \& {Gal-Yam}, A. 2017,
  \JournalTitle{ArXiv e-prints},
  \href{http://arxiv.org/abs/1711.09638}{{\sffamily arXiv:1711.09638
  [astro-ph.HE]}}

\bibitem[{{Wollaeger} {et~al.}(2017){Wollaeger}, {Korobkin}, {Fontes},
  {Rosswog}, {Even}, {Fryer}, {Sollerman}, {Hungerford}, {van Rossum}, \&
  {Wollaber}}]{Wollaeger+17}
{Wollaeger}, R.~T., {Korobkin}, O., {Fontes}, C.~J., {et~al.} 2017,
  \JournalTitle{ArXiv e-prints},
  \href{http://arxiv.org/abs/1705.07084}{{\sffamily arXiv:1705.07084
  [astro-ph.HE]}}

\bibitem[{{Wu} {et~al.}(2016){Wu}, {Fern{\'a}ndez}, {Mart{\'{\i}}nez-Pinedo},
  \& {Metzger}}]{Wu+16}
{Wu}, M.-R., {Fern{\'a}ndez}, R., {Mart{\'{\i}}nez-Pinedo}, G., \& {Metzger},
  B.~D. 2016, \href{http://dx.doi.org/10.1093/mnras/stw2156}{\JournalTitle{Mon.
  Not. R. Astron. Soc.}}, \href{http://arxiv.org/abs/1607.05290}{{\sffamily
  arXiv:1607.05290 [astro-ph.HE]}}

\bibitem[{{Zrake} \& {MacFadyen}(2013)}]{Zrake&MacFadyen13}
{Zrake}, J., \& {MacFadyen}, A.~I. 2013,
  \href{http://dx.doi.org/10.1088/2041-8205/769/2/L29}{\JournalTitle{Astrophys.
  J. Lett.}, 769, L29}

\end{thebibliography}

\end{document}